\documentclass{statsoc}

\usepackage{bm}
\usepackage{amsmath, amssymb,fullpage}
\usepackage{graphicx}
\usepackage[margin=1in]{geometry}
\usepackage{xcolor}
\usepackage{algpseudocode}
\usepackage{verbatim}
\usepackage[utf8]{inputenc}
\usepackage{url}
\usepackage{natbib}
\newcommand{\beginsupplement}{%
        \setcounter{table}{0}
        \renewcommand{\thetable}{S\arabic{table}}%
        \setcounter{figure}{0}
        \renewcommand{\thefigure}{S\arabic{figure}}%
}

\newtheorem{theorem}{Theorem}
\newtheorem{remark}{Remark}
\newtheorem{proposition}{Proposition}
 
\title{A Bayesian hierarchical model for related densities using P\'olya trees}
\author{Jonathan Christensen}
\address{Duke University,
Durham, NC, USA.}
\email{jonathan.christensen@duke.edu}
\author[Christensen and Ma]{Li Ma}
\address{Duke University,
Durham, NC, USA.}
\email{li.ma@duke.edu}

\begin{document}
\maketitle

\begin{abstract}
Bayesian hierarchical models are used to share information between related samples and obtain more accurate estimates of sample-level parameters, common structure, and variation between samples.
When the parameter of interest is the distribution or density of a continuous variable, a hierarchical model for continuous distributions is required.
A number of such models have been described in the literature using extensions of the Dirichlet process and related processes, typically as a distribution on the parameters of a mixing kernel.
We propose a new hierarchical model based on the P\'olya tree, which allows direct modeling of densities and enjoys some computational advantages over the Dirichlet process.
The P\'olya tree also allows more flexible modeling of the variation between samples, providing more informed shrinkage and permitting posterior inference on the dispersion function, which quantifies the variation among sample densities.
We also show how the model can be extended to cluster samples in situations where the observed samples are believed to have been drawn from several latent populations.
\end{abstract}

\section{Introduction}

Many statistical applications deal with learning and comparing the distributions of two or more related samples.
We may be interested in learning how samples are similar or testing whether they are distinguishably different from each other.
Because distributions are complex infinite-dimensional objects, classical approaches work with low-dimensional parameterizations of the distribution.
Analysis of variance, for example, reduces distributions to a mean and variance, which are sufficient under the assumption of normality.
A wide range of other parametric models within both the Bayesian and frequentist inferential frameworks use other parameterizations of the distribution to reduce the dimensionality of the problem.
Classical nonparametric approaches use features of the samples such as medians \citep{Westenberg1948}, rank-based scores \citep{Wilcoxon1945}, or summaries of the empirical distribution functions, as in the Kolmogorov-Smirnov \citep{Kolmogorov1933} and Cram\'{e}r-von Mises tests \citep{Anderson1962}.

A number of Bayesian nonparametric approaches embrace the infinite-dimensional nature of the problem using extensions of Dirichlet processes \\citep{ferguson1973}.
Among the most well-known, The Hierarchical Dirichlet process \citep{beal2002hdp,hdp} builds a hierarchical model using the Dirichlet process, allowing it to share information between samples.
\cite{Tomlinson1999} provide an early description of a similar model.
The Nested Dirichlet process of \cite{ndp} takes a different approach, using a Dirichlet process as the base measure of a second Dirichlet process. This induces clustering in the samples, with samples in a cluster being modeled with a single density.
\cite{hdpm} model each sample density as a mixture of two components: one Dirichlet process mixture of Gaussians representing common structure between samples, and a second Dirichlet process mixture of Gaussians representing the idiosyncratic structure of the given sample.
A variety of other dependent Dirichlet Processes \citep{maceachern1999dependent} have been described in the literature. \cite{Teh_Jordan_2010} give an overview of hierarchical models based on the Dirichlet process.

While the Dirichlet process has been the basis of most of the work in this area, work has also been done on hierarchical extensions of other priors. For example, \cite{Teh2006a} defines a Hierarchical Pitman-Yor process, taking advantage of the more flexible clustering structure of the Pitman-Yor process over the Dirichlet process. \cite{Camerlenghi2017} consider hierarchical models based on Normalized Random Measures \citep{Barrios2013}, which includes the Dirichlet process as a special case. 

An alternative approach for modeling densities within the Bayesian nonparametric framework is to use a model derived from P\'olya trees \citep{freedman1963}.
P\'olya trees are a class of tail-free prior in which an infinite recursive binary partition is placed on the sample space, and probability mass assigned to the elements of the partition by a corresponding infinite sequence of Beta-distributed random variables.
A special case gives the familiar Dirichlet process \citep{ferguson1973}, but the P\'olya tree family is considerably more flexible.
With appropriate specification of the prior parameters, the P\'olya tree assigns probability 1 to the set of absolutely continuous distributions \citep{Kraft1964,ferguson1974}.
This property allows it to be used to directly model probability densities without the encumbrance of a mixing kernel.

The P\'olya tree model allows tractable computation of the marginal likelihood, which has made it popular in Bayesian hypothesis testing of nonparametric density models.
Examples include \cite{berger2001, ma2011, chen&hanson2014, holmes2015, soriano&ma2017, filippi2017}.
These approaches are unsatisfactory when estimation and prediction rather than formal hypothesis testing are of primary interest.
Statisticians have recognized the benefits of partial shrinkage as far back as Stein's shrinkage estimator for the mean of a multivariate normal distribution \citep{stein1956}.
We may expect that related samples will have similar but not identical distributions.
In this situation, neither independence nor a single common distribution are appropriate models.
A partial shrinkage model allows borrowing of information across samples while preserving cross-sample variation.
The Polya tree has also been used as a building block for modeling dependent densities. \cite{zhao2011} describe a spatially dependent Polya tree with an autoregressive structure.
When the spatial dependence is removed, a partially exchangeable model similar to the one we consider here arises.
 \cite{schorgendorfer2013} and \cite{nieto2016bayesian} describe dependent Polya tree models with autoregressive structures inducing dependence among densities over time. \cite{JaraHanson} transform Gaussian processes in the framework of tail-free processes to create a regression model for dependent densities.

Within the Bayesian inference framework, partial shrinkage is naturally achieved with a hierarchical model.
While a simple hierarchy of P\'olya trees is possible and is described herein, we build on the richer Adaptive P\'olya tree model of \cite{apt}.
This model allows us to learn the concentration parameters of the P\'olya tree, rather than fixing them in the prior.
In contrast to Dirichlet process-based models, the richer structure of the P\'olya tree allows us to model both the means and the cross-sample variation of the densities in a fully nonparametric manner.
This not only improves the estimation of the distributions, but also allows us to perform inference on how the variation across samples differs over the sample space.

In Section~\ref{model} we describe the model and contrast it with several existing models.
We discuss Bayesian posterior inference and computation in Section~\ref{BayesianInference}.
In Section~\ref{theory} we give theoretical grounding for the appropriateness of the method.
Section~\ref{methodological-applications} describes two ways in which the model allows us to go beyond the capabilities of existing models.
We show simulation results in Section~\ref{simulation} and demonstrate application to real data in Section~\ref{applications}.
Finally, we conclude with some discussion in Section~\ref{discussion}.

\section{Model}
\label{model}

\subsection{Reviewing the P\'olya tree construction}

\begin{center}
\begin{figure}
\includegraphics[width=\textwidth]{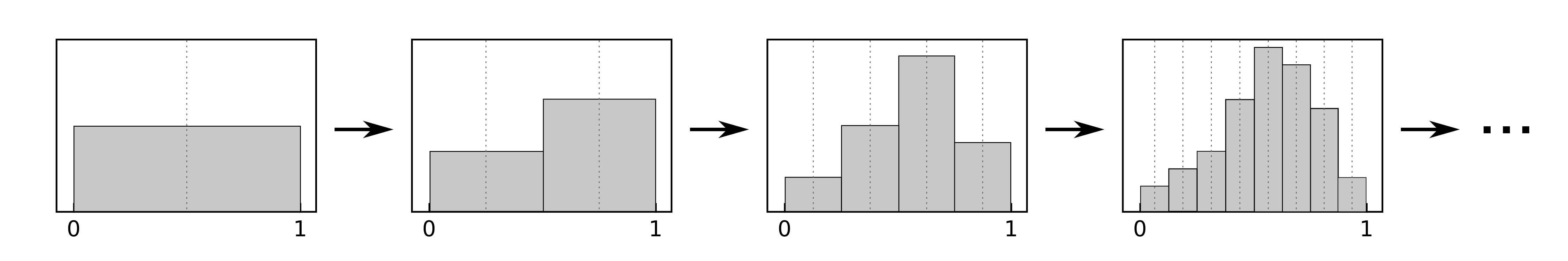}
\caption{An illustration of the recursive partitioning and probability allocation of the standard P\'olya tree.}
\label{polyatree}
\end{figure}
\end{center}

We begin with a brief sketch of the P\'olya tree; the reader interested in the mathematical details should refer to \cite{mauldin1992, lavine1992, lavine1994}. 
The P\'olya tree consists of an infinite recursive partition $\mathcal{A}$ of the sample space and a corresponding infinite sequence of Beta-distributed random variables which assign mass to the various regions $A \in \mathcal{A}$ of the partition.
Figure~\ref{polyatree} illustrates the partitioning sequentially.
In this illustration our sample space is the interval $(0,1]$ on the real line, and our prior mean, shown in the first pane, is the uniform distribution on that interval.
At each level of the recursive partition we cut each region in half at the midpoint.
Although arbitrary partitions may be used, the dyadic partition described here is convenient and is often used as a default partition in the absence of a compelling reason to use a different one. Other partitions may be more convenient in the presence, for example, of censored data \citep{Muliere1997}.
The second pane shows the result after the first cut and mass allocation, in this case the majority of the mass having been allocated to the right-hand side.
In the next step we cut each of the two regions from the second pane in half again, and assign the probability mass to each to its children according to a Beta-distributed random variable.
This results in four regions, shown in the third pane, which are again cut and mass distributed in the fourth pane.
The process continues indefinitely.

We denote the P\'olya tree prior as $Q \sim \text{PT}(Q_0, \bm\nu)$, where $Q_0$ is the centering distribution and $\bm\nu$ is an (infinite-dimensional) concentration parameter describing how much $Q$ is expected to vary from $Q_0$. The parameters of the sequence of Beta distributions from which the mass allocations are drawn are derived very simply from $Q_0$ and $\bm\nu$.
For an arbitrary region $A$ belonging to the recursive partition $\mathcal{A}$, the fraction of the mass allocated to the left child $A_\ell$ of $A$ is given by the random variable
\[\theta(A) \sim \text{Beta}(\theta_0(A)\nu(A), (1-\theta_0(A))\nu(A)),\]
where $\theta_0(A) = Q_0(A_\ell) / Q_0(A),$
with the remainder of the mass allocated to the right child $A_r$.
$Q_0$ thus determines the mean of the mass allocations (and hence the expectation of the resulting density), while $\bm\nu = \{\nu(A): A \in \mathcal{A}\}$ controls the variation of the mass allocations, and hence the dispersion of $Q$ around $Q_0$. Alternatively, $\bm\nu$ controls the strength of the shrinkage of the posterior mean density from the empirical process towards $Q_0$.
With an appropriate choice of $\bm\nu$, the P\'olya tree prior almost surely generates an absolutely continuous distribution \citep{Kraft1964}.

\subsection{The Hierarchical P\'olya Tree}

\begin{center}
\begin{figure}[t]
\includegraphics[width=\textwidth]{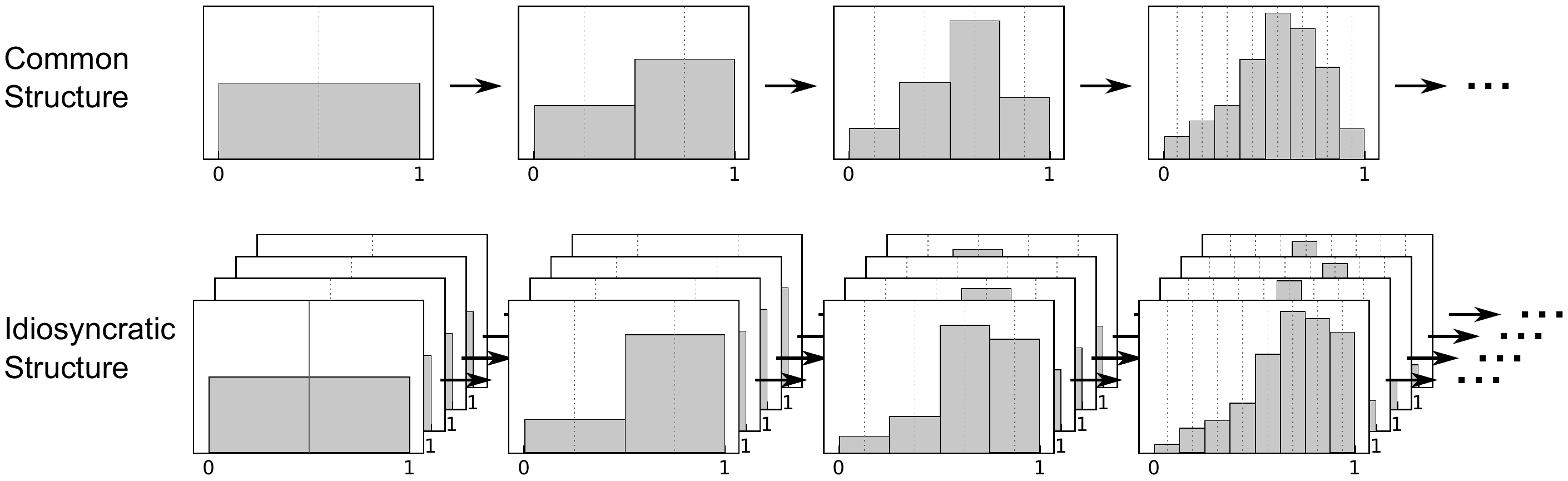}
\caption{An illustration of the Hierarchical P\'olya tree.}
\label{hpt}
\end{figure}
\end{center}

It is conceptually straightforward to extend the P\'olya tree to a hierarchical model. Let $\mathbf{X}_1, \dotsc, \mathbf{X}_k$ be $k$ samples arising from related distributions on a complete, separable space $\Omega$.
For ease of exposition we again use $\Omega = (0,1]$, though like the P\'olya tree the model can be applied to more general sample spaces.
We model these samples as coming from $k$ exchangeable distributions $Q_i$, which are centered at a common underlying measure $Q$, itself unknown.
Applying P\'olya tree priors to both $Q$ and the $Q_i$, all with identical partition structures, gives us the hierarchical model
\begin{equation}
\label{HPTmodel}
\begin{split}
X_{ij} \mid Q_i &\overset{ind}{\sim} Q_i \\
Q_i \mid Q &\overset{iid}{\sim} \text{PT}(Q, \bm\tau) \\
Q &\overset{\phantom{iid}}{\sim} \text{PT}(Q_0, \bm\nu).
\end{split}
\end{equation}
Here $Q_0$ is the overall prior mean, $\bm\tau$ controls the variation across samples around the common structure $Q$, and $\bm\nu$ controls the variation of the common structure from $Q_0$, which determines the smoothness of $Q$.
This model, which we call the Hierarchical P\'olya tree, allows nonparametric estimation of the sample distributions $Q_i$ and the common structure $Q$.
The Hierarchical P\'olya tree model is illustrated in Figure~\ref{hpt}.
The first row shows the upper level of the hierarchy, which like the basic P\'olya tree illustrated in Figure~\ref{polyatree} is centered at the uniform distribution on $(0,1]$.
The second row shows the lower level of the hierarchy, the individual sample distributions $Q_i$ conditioned on $Q$.
They follow exactly the same P\'olya tree construction, but each cut is centered on the corresponding cut from $Q$, rather than on the uniform distribution.
$Q$ captures the common structure of the samples, while the deviation of $Q_i$ from $Q$ captures the idiosyncratic structure of each sample.
Because the partition structures are identical, the hierarchy of P\'olya trees translates directly to the decomposed space as a hierarchical model for Beta random variables. For an arbitrary region $A \in \mathcal{A}$, we have
\begin{equation}
\begin{aligned}
\theta_i(A) \mid \theta(A) &\overset{iid}{\sim} \text{Beta}(\theta(A)\tau(A), (1-\theta(A))\tau(A))\\
\theta(A) &\sim \text{Beta}(\theta_0(A)\nu(A), (1-\theta_0(A))\nu(A)).
\end{aligned}
\label{hierarchical-beta}
\end{equation}
The representation of the hierarchy of P\'olya trees as a hierarchy of Beta random variables allows tractable posterior inference, as described in Section~\ref{BayesianInference}.

\subsection{The Stochastically Increasing Shrinkage prior on dispersion}

The P\'olya tree's concentration parameter is traditionally set to increase with depth at a predetermined rate to ensure absolute continuity, with a constant multiplicative term to control the overall level of variation which may be treated as a tuning parameter (as in \cite{berger2001}) or have a prior placed on it.
\cite{Hanson2006} discusses some of the necessary considerations when placing a prior on this parameter.
More recently, \cite{apt} shows that putting a flexible nonparametric prior on the concentration parameter allows the P\'olya tree to learn the true distribution more accurately, particularly when the smoothness of the underlying density varies over the sample space.
We can extend the Hierarchical P\'olya tree model by placing priors on both concentration parameters $\bm\tau$ and $\bm\nu$.
In addition to more accurate inference on $Q$ and the $Q_i$, putting a prior distribution on $\bm\tau$ allows us to learn the variation across samples in a nonparametric way.
That is, we can estimate a posterior \emph{dispersion function} which summarizes the variability across sample densities at any given point in the sample space.
Dispersion can be measured in a variety of ways; in Section~\ref{variation} we show how to derive the posterior mean variance of the densities and interpret a standardized version using the coefficient of variation to correct for the height of the density.
Nonparametric inference on the variation across samples over the sample space is made possible by the flexibility of the P\'olya tree model.
We contrast how several other models treat cross-sample variation in Section~\ref{compare}.

A simple approach is to put independent priors on the variance of each Beta random variable.
However, we expect spatial structure in the dispersion function---locations near each other in the sample space are likely to have similar levels of variation. The variance of the Beta distributions is generally expected to be smaller at deeper levels of the partition, but the decay in the variance may be heterogeneous over the sample space depending on the local smoothness of the densities.
While the recursive partitioning allows independent priors to capture some spatial structure, we can do better by introducing dependency between regions in the partition.
\cite{apt} introduces Markov dependency on the concentration parameter, following the tree topology. The Markov dependency among the Beta variance parameters allowa the Beta variables to stochastically transition into lower prior variance (i.e., higher shrinkage) along each branch of the partition tree, but at potentially different rates, thereby allowing spatially heterogeneous prior variability in the random densities. Instead of constructing a continuous state-space model on the prior Beta variance directly, which would make posterior computation very challenging, a discrete latent state variable that characterizes a finite number of different levels of prior variance is introduced for each Beta variable, and the Markov dependency is then imposed on these latent state variables. This discretization strategy maintains the computational tractability of the P\'olya tree model when equipped with more flexible prior on the variance parameters.

\begin{figure}[t]
\begin{center}
\includegraphics[trim={2em 5em 0 15em}, clip=TRUE, width=1\textwidth]{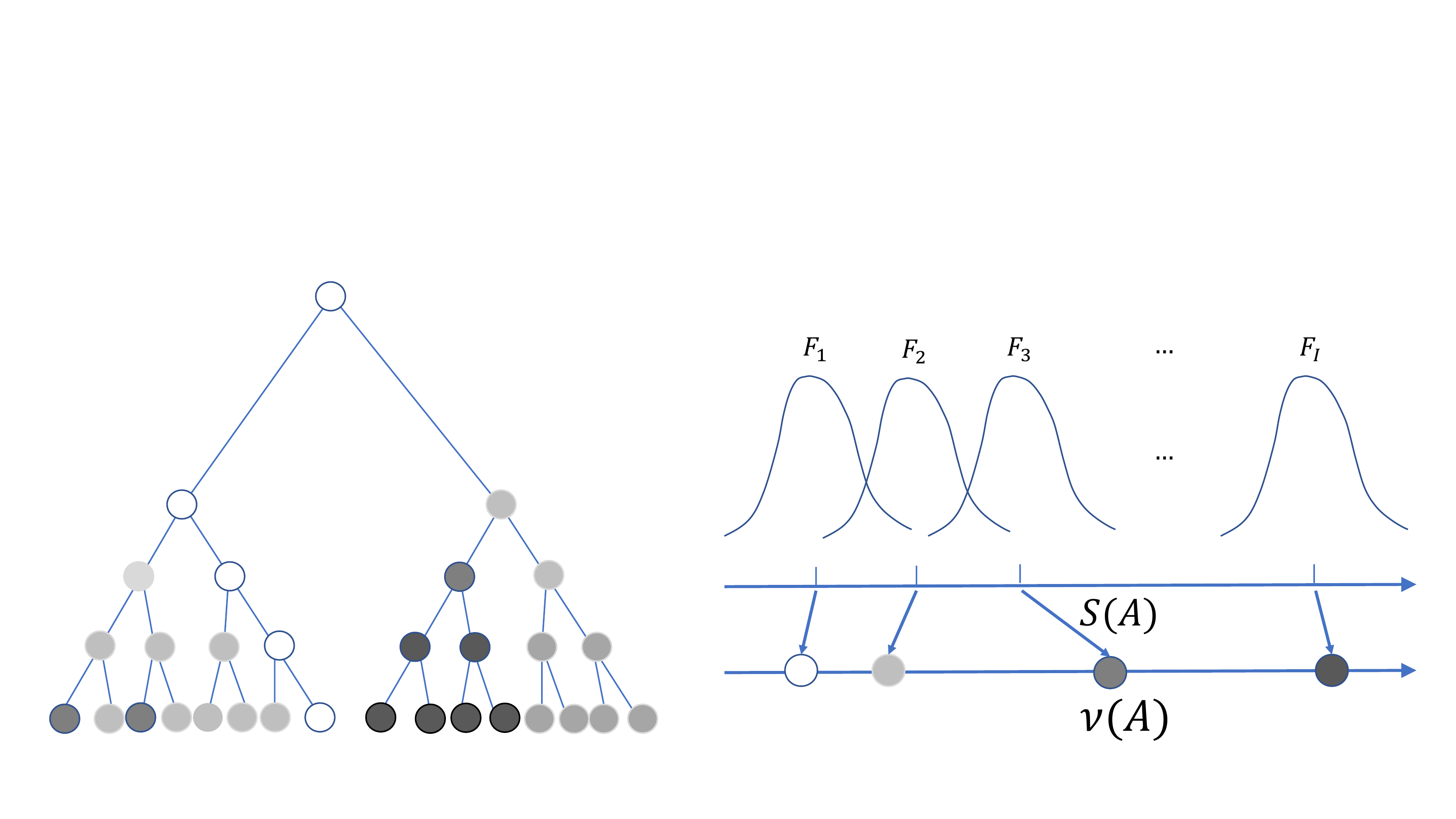}
\end{center}
\caption{Illustration of the SIS prior. The shrinkage states increase as you follow the tree down to finer scales---indicated by darker shades of gray---at potentially different rates across the space, eventually reaching complete shrinkage---but allowing less shrinkage where the data dictates such.}
\label{SIS}
\end{figure}

More specifically, the Stochastically Increasing Shrinkage (SIS) prior introduces a state variable $S(A)$ supported on a finite set of integers $1, \dotsc, I$, corresponding to decreasing prior variance and increasing shrinkage for the Beta random variables. In particular, the last state $I$ corresponds to complete shrinkage or zero variance, which is achieved through fixing $\nu(A)$ at $\infty$ when $S(A)=I$. 
For example, we may have $S(A) \in \{1, 2, 3, 4\}$ with $S(A) = i$ implying $\nu(A)\,|\,S(A)=i \sim F_i$ with the $F_i$ stochastically ordered $F_1 \prec F_2 \prec F_3 \prec F_4$ and $F_4$ being a point mass at infinity, corresponding to zero variance.
The number of states and the corresponding distributions can be chosen to balance the flexibility and computational complexity of the model.
A simple way to enforce such a stochastic ordering is through partitioning the support of the concentration parameter $\nu(A)$ into disjoint intervals. 
Given these latent states, the SIS prior adopts a transition probability matrix $\Gamma(A)$ for $S(A) \mid S(\text{Par}(A))$, where $\text{Par}(A)$ represents the parent node of $A$ in the tree.
\cite{apt} discusses several prior possibilities for this transition matrix. We adopt the recommendation given there to use an exponential kernel for the transition probabilities.
The resulting transition matrix can be given as
\[\Gamma(A) = 
\begin{bmatrix}
\frac{1}{\sum_{i=0}^{I-1} e^{\beta \cdot i}} & \frac{e^{\beta}}{\sum_{i=0}^{I-1} e^{\beta \cdot i}} & \dotsm & \frac{e^{\beta (I-1)}}{\sum_{i=0}^{I-1} e^{\beta \cdot i}}\\[1em]
0 & \frac{1}{\sum_{i=0}^{I-2} e^{\beta \cdot i}} & \dotsm & \frac{e^{\beta (I-2)}}{\sum_{i=0}^{I-2} e^{\beta \cdot i}}\\
\vdots & \vdots & \ddots & \vdots \\
0 & 0 & \dotsm & 1 \\
\end{bmatrix}.\]
This upper-triangular transition probability matrix induces stochastically increasing shrinkage (or decreasing prior Beta variance) along each branch in the partition tree (see Figure~\ref{SIS}), and ensures that the model generates absolutely continuous densities (see Theorem~\ref{thm:abs_continuity}).
The parameter $\beta$ controls the stickiness of the transition; that is, larger values of beta correspond to a stronger dependence in the shrinkage state between adjacent nodes in the P\'olya Tree.
$\beta$ can be set equal to zero, in which case the transition is uniform over the states with shrinkage no less than the current node.
We denote the SIS prior
\[\bm\nu \sim \text{SIS}(\bm\Gamma).\]

Figure~\ref{SIS} illustrates the SIS prior in action. The gray-scale in each node indicates the value of the Beta variance, with darker gray indicates less variance and higher shrinkage. These Beta variances are determined by a set of latent state variables, each corresponding to a conditional prior $F_i$ on $\nu(A)$, represented in the right panel of the figure. As you move down to finer resolutions the shrinkage state tends to increase, eventually reaching complete shrinkage, indicated by black nodes, though the rates at which the shrinkage increases along different branches of the tree are different and are determined stochastically by the underlying top-down Markov model linking the latent state variables. This model allows the amount of shrinkage or prior variance to vary over the sample space to capture large-scale smooth features in one part of the sample space and smaller scale features in another part. This allows the resulting density to have heterogeneous smoothness and variability around the mean across the sample space.

\subsection{The Hierarchical Adaptive P\'olya Tree}

\begin{figure}
\begin{center}
\includegraphics[width=4.5in]{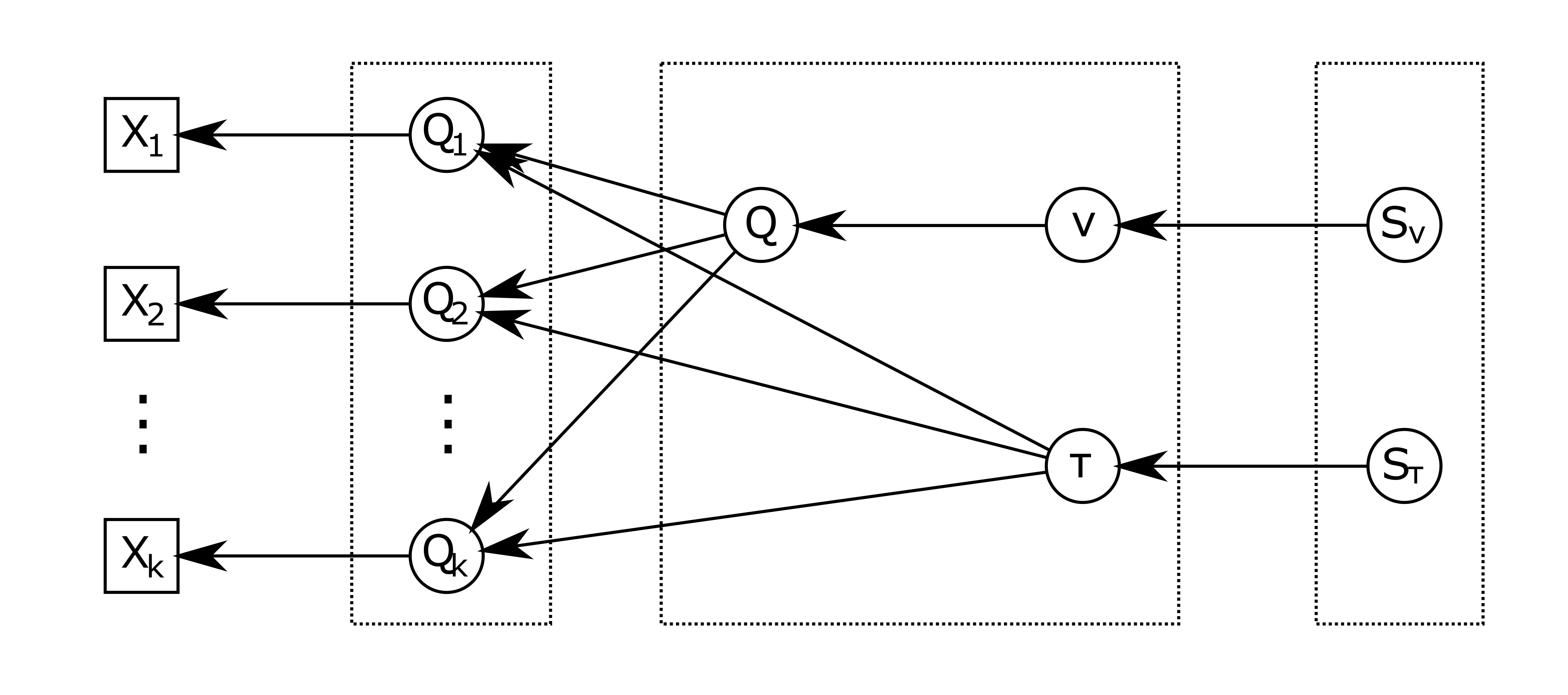}
\caption{
A graphical representation of the HAPT model.
The boxes outline the parts of the model whose posteriors are computed with each of the three strategies described in Section~\ref{hapt-computation}.
From left to right: The posterior of the $Q_i$ conditional on other parameters is conjugate and can be integrated out numerically; the posterior of $Q, \bm\nu,$ and $\bm\tau$ conditional on $\bm{S_\nu}$ and $\bm{S_\tau}$ is approximated using quadrature; and the posterior of $\bm{S_\nu}$ and $\bm{S_\tau}$ is computed using HMM methods.
}
\label{modelgraph}
\end{center}
\end{figure}

Having described the SIS prior, we can adopt this prior for the concentration parameters $\bm\tau$ and $\bm\nu$, and write the complete model as follows:
\begin{align*}
X_{ij} \mid Q_i &\overset{ind}{\sim} Q_i \\
Q_i \mid Q, \bm\tau &\overset{iid}{\sim} \text{PT}(Q, \bm\tau) \\
Q \mid \bm\nu &\overset{\phantom{iid}}{\sim} \text{PT}(Q_0, \bm\nu) \\
\bm\tau &\overset{\phantom{iid}}{\sim} \text{SIS}(\bm\Gamma_{\bm\tau}) \\
\bm\nu &\overset{\phantom{iid}}{\sim} \text{SIS}(\bm\Gamma_{\bm\nu}).
\end{align*}
We call this model the Hierarchical Adaptive P\'olya Tree (HAPT).
In contrast to existing models, this specification allows fully nonparametric inference on both the densities and the variation across densities.
The inclusion of the SIS priors allows the model to adapt the level of shrinkage or information borrowing in different parts of the sample space to more accurately capture the density of each sample, rather than using fixed uniform shrinkage.
Figure~\ref{modelgraph} gives a graphical representation of the model, showing the conditional relationships between parameters. The three boxes indicate the sets of parameters that are handled using three different computational strategies in the calculation of the posterior; see Section~\ref{hapt-computation} for details.

\subsubsection{Choice of prior parameters}

The Hierarchical Adaptive P\'olya Tree has three prior parameters for which values must be chosen. The most straightforward is the prior mean measure $Q_0$, which we typically chose from a parametric family. Some consideration must be given to the choice of the SIS prior parameters $\bm\Gamma_\tau$ and $\bm\Gamma_\nu$.

\cite{apt} recommends an empirical Bayes approach to setting the $\beta$ parameters, and where computationally feasible we may also recommend this approach.
When the dataset is quite large or HAPT is embedded in a larger algorithm, as in Section~\ref{hapt-dp-section}, empirical Bayes estimation may not be practical.
Based on our experience the model fit is not very sensitive to reasonable variation in this parameter; we suggest default values of $\beta = 1$ for both $\bm\Gamma_\tau$ and $\bm\Gamma_\nu$.

\subsection{Comparison to existing models}
\label{compare}

The most prominent existing nonparametric models for estimation of related distributions are based on the Dirichlet process, including the Hierarchical Dirichlet process (HDP) \citep{hdp}, Nested Dirichlet process (NDP) \citep{ndp,rodriguez2014}, and the hierarchical mixture of common and idiosyncratic Dirichlet process model of M\"uller, Quintana, and Rosner (MQR) \citep{hdpm}.
The Hierarchical Adaptive P\'olya tree enjoys several advantages over these methods:

\begin{enumerate}
\item \textbf{Nonparametric estimation of cross-sample variation.}
The HDP and NDP have a single scalar concentration parameter that controls the dependence across samples.
MQR has one scalar parameter per sample that controls what proportion of the sample is explained by common structure and how much by idiosyncratic structure.

In contrast, because the concentration parameter $\bm\tau=\{\tau(A):A\in \mathcal{A}\}$ in the HAPT is infinite-dimensional, HAPT places a highly flexible prior on the variation across samples, which allows it to learn spatially heterogeneous variation across samples. 
Indeed, the variation between samples at different locations of the sample space may be of primary interest in some scientific applications: learning where common structure is largely preserved between samples and where distributions vary widely may point the way to understanding important underlying phenomena.
\vspace{0.25em}

\item \textbf{Computation.} HDP, NDP, and MQR all rely on MCMC methods to draw from the posterior. The Hierarchical Adaptive P\'olya tree is not fully conjugate, but the necessary integration can be split into low-dimensional integrals and approximated extremely quickly using adaptive quadrature methods, without concerns about Markov chain convergence. See Section~\ref{hapt-computation} for details. Note that this is not to say that one will never use MCMC in the presence of HAPT in any Bayesian model. In some inference tasks, one may embed HAPT into a more complex hierarchical model, whose other components may require MCMC for inference. In such cases, the computational tractability of HAPT implies that one can Rao-Blackwellize (i.e., marginalize out) the HAPT part within that MCMC algorithm for the more complex model. This will substantially simplify the MCMC sampler. We present one such application of HAPT in Section~\ref{hapt-dp-section}.
\vspace{0.25em}

\item \textbf{Interpretability.}
HAPT provides an easily interpretable estimate of the common structure: The posterior estimate of $Q$ is both the estimate of the mean density across samples, and the posterior predictive distribution for a new sample.
In contrast, the HDP estimates a discrete instead of continuous distribution, and the NDP does not provide any estimate of common structure.
MQR provides an estimate of common structure, but it is neither the mean of sample distributions nor a posterior predictive estimate.
Interpretation of the common structure in the MQR model is most straightforward if variation between samples involves contamination of an underlying distribution by an idiosyncratic process for each sample.
\end{enumerate}

Other existing models such as the Hierarchical Pitman-Yor model \citep{Teh2006a} and those based on Normalized Random Measures \citep{regazzini2003} or more generally Poisson-Kingman models \citep{pitman2003} are subject, to various extents, to limitations similar to those of the models based on the Dirichlet process. Recent works such as \cite{griffin2013,griffin2017,Camerlenghi2017,camerlenghi2018} provide more flexible means to modeling the dependency among multiple samples than the classical models, but the aforementioned benefits of the P\'olya tree remain even in view of these state-of-the-art models.

\section{Bayesian inference and computation}\label{BayesianInference}

The HAPT model is partially conjugate: the conditional posterior for $Q_i \mid Q, \bm\tau, \bm\nu, \bm{S}_{\bm\tau}, \bm{S}_{\bm\nu}$ is a standard P\'olya tree. Though not fully conjugate, the conditional posterior for $Q, \bm\tau, \bm\nu \mid \bm{S}_{\bm\tau}, \bm{S}_{\bm\nu}$ can be reliably approximated using adaptive quadrature methods. The computational strategies used are described in Section~\ref{hapt-computation}

To derive the posterior we use the representation of the P\'olya trees $Q$ and $Q_i$ in terms of Beta-distributed random variables $\theta(A)$ and $\theta_i(A)$ for each node $A$ of the tree. With this notation, The second and third lines in (\ref{HPTmodel}) can be written in terms of the $\theta$ and $\theta_i$ as in Equation \ref{hierarchical-beta}:
\begin{align*}
\theta_i(A) \mid \theta(A) &\overset{iid}{\sim} \text{Beta}\left(\theta(A)\tau(A), (1-\theta(A))\tau(A)\right)\\
\theta(A) &\overset{\phantom{iid}}{\sim} \text{Beta}\left(\theta_0(A)\nu(A), (1-\theta_0(A))\nu(A)\right).
\end{align*}

Including the concentration parameters, we can write the posterior for the parameters of a region $A$ in the following form conditional on the state parameters $S_\tau(A), S_\nu(A)$:
\begin{equation}
\begin{aligned}
\pi(\theta(A), \tau(A), \nu(A) &\mid S_\tau(A), S_\nu(A), \bm{X}) \propto \\
& \theta(A) ^{\theta_0(A)\nu(A) -1} (1-\theta(A)) ^{(1-\theta_0(A))\nu(A) -1}\times \\
& \left[B\left(\theta(A)\tau(A), (1-\theta(A))\tau(A)\right)\right]^{-k} \times \\
& \prod_{i=1}^k B(\theta(A)\tau(A) + n_i(A_\ell), (1-\theta(A))\tau(A) + n_i(A_r)) \times\\
& \pi(\tau(A) \mid S_\tau(A))\pi(\nu(A) \mid S_\nu(A)).
\end{aligned}
\label{node-posterior}
\end{equation}
where $B(\cdot, \cdot)$ is the Beta function and the function $n_i(\cdot)$ counts the number of observations from the $i$th sample contained in a region.
The full posterior is the summation of Equation (\ref{node-posterior}) over the possible states of $\bm{S_\tau}$ and $\bm{S_\nu}$, with their respective priors factored in. The derivation of this posterior is given in the supplementary material.

\subsection{Computation}
\label{hapt-computation}

Posterior computation of the HAPT model requires three distinct computational strategies.
We split the model (see Figure~\ref{modelgraph}) into three parts, each of which requires a different approach.
Each part of the model is conditioned on all parameters which are further to the right in Figure~\ref{modelgraph}.
We first describe how to integrate out each of the first two parts. This reduces the problem to evaluating the posterior probabilities of all possible combinations of the state variables $\bm{S}_{\bm\tau}, \bm{S}_{\bm\nu}$, which can be accomplished using a forward-backward algorithm for hidden Markov models.
\begin{enumerate}
\item $\pi(Q_i \mid Q, \bm\tau, \bm\nu, \bm{S}_{\bm\tau}, \bm{S}_{\bm\nu}, \bm{X})$: The individual sample densities $Q_i$, conditional on all other parameters, are \emph{a priori} distributed according to a standard P\'olya tree.
The corresponding conditional posterior is therefore also a P\'olya tree.
This allows us to analytically integrate out the $Q_i$ when computing the posterior.
If individual sample densities are of inferential interest their posteriors can easily be reconstructed after the main posterior computation is completed.

\item $\pi(Q, \bm\tau, \bm\nu \mid \bm{S}_{\bm\tau}, \bm{S}_{\bm\nu}, \bm{X})$: The remaining continuous parts of the joint model, namely the common structure $Q$ and the continuous concentration parameters $\bm\tau$ and $\bm\nu$, conditioned on the discrete state parameters $\bm{S}_{\bm\tau}$ and $\bm{S}_{\bm\nu}$, are not conjugate and must be integrated numerically.
Because all parameter dependence across nodes in the P\'olya tree topology is restricted to the discrete state parameters, by conditioning on those parameters we are able to compute the posterior of the remaining parameters for each node of the tree independently.

This has two significant implications.
First, rather than tackling a very high-dimensional integral over the product space of the parameters for all nodes, we have a much more tractable collection of low-dimensional integrals: we need only integrate the three-dimensional joint posterior of $\theta(A), \tau(A), \nu(A)$ for each region $A$ in the P\'olya tree.
Each of these integrals is tractable using standard quadrature techniques.
Second, these integrals can be computed in parallel.

An additional observation allows us to further accelerate the adaptive quadrature.
We note that the joint posterior distribution for $\theta(A), \tau(A), \nu(A)$ conditional on $S_\tau(A)$ and $S_\nu(A)$, with the other parameters integrated out, can be factored as
\[\pi(\theta(A), \tau(A), \nu(A) \mid S_\nu(A), S_\tau(A), \bm{X}) = g(\theta(A), \tau(A)) \times h(\theta(A), \nu(A)).\]
This allows us to factor the three dimensional integral:
\begin{align*}
&\iiint \pi(\theta(A), \tau(A), \nu(A) \mid S_\nu(A), S_\tau(A), x) \; d\tau(A)\, d\nu(A)\, d\theta(A) \\
&\qquad\qquad= \int \left[\int g(\theta(A), \tau(A)) \; d\tau(A)\right] \left[\int h(\theta(A), \nu(A)) \; d\nu(A)\right] d\theta(A),
\end{align*}
This factorization effectively reduces the dimensionality of the integral: rather than evaluating the unnormalized posterior at points throughout a 3-dimensional space, we need only evaluate it on the union of two 2-dimensional spaces.

\item The posterior distribution of the state parameters $\bm{S_\tau}$ and $\bm{S_\nu}$ at first appears to be the most intimidating part of the model: It is a distribution over the product space of a large number of discrete parameters, resulting in an enormous number of level combinations. The naive computation of the joint posterior,
\[P\left(\bigcap_{A \in \mathcal A} S_\tau(A) = i_A, S_\nu(A) = j_A\right)\]
is straightforward but needs to be repeated for every possible combination of $S_\tau(A)$ and $S_\nu(A)$ for every node in the tree, which is computationally prohibitive. Here the Markov dependency structure comes to our rescue. The shrinkage states constitute a Hidden Markov Model following the tree topology \citep{Crouse1998}, and we can factor the joint distribution and calculate the posterior probabilities using a forward-backward algorithm in a manner analogous to inference strategies for linear Hidden Markov Models. 

During the forward-backward algorithm we can compute expectations of any function that can be expressed in the form
\[f(\cdot) = \prod_{A} f^*(\theta(A), \tau(A), \nu(A)),\]
where $f^*(\cdot)$ is an arbitrary function in $L_1$.
This includes the marginal likelihood, the expected value of the estimated common density $q(\cdot)$ or any individual sample density $q_i(\cdot)$  at any given point, expectations of random variables $Y \sim Q$ or $Y_i \sim Q_i$, and a wide variety of other functions, such as the variance function described in Section~\ref{variation}.

This computation is recursive, and we give a brief example of how it is carried out for the marginal likelihood. The previous two computational strategies give us the ability to calculate (up to quadrature approximation) the marginal likelihood, within a given node, of all remaining parameters conditional on $\bm{S_\tau}$ and $\bm{S_\nu}$. Combining this with the prior transition parameters specified in $\bm\Gamma$ we are able to calculate the posterior probabilities for $S_\tau(A)$ and $S_\nu(A)$, and the overall marginal likelihood of the distribution on $A$, by considering the recursively-calculated marginal likelihoods of the child regions of $A$ under each possible state.

Obviously this recursion requires a stopping point. The simplest method is to truncate the tree at a predetermined depth.
\cite{Hanson2006} offers some guidance on how to choose the depth of a truncated P\'olya tree based on sample size and other considerations.
We recommend using as large a tree as is computationally feasible in order to minimize approximation errors due to truncation. 
If the data deviate very strongly from the prior distribution---as is common, for example, in high-dimensional settings---a more sophisticated approach may be required, such as truncating a branch of the tree when it reaches a depth where the node contains only a few data points.
\end{enumerate}

\section{Theoretical results}
\label{theory}
We describe several desirable theoretical properties of the HAPT model. Proofs are given in the supplementary material.
\begin{theorem} (Absolute Continuity)
\label{thm:abs_continuity}
Let $Q, Q_1, \dotsc, Q_k$ be random measures distributed according to a HAPT model with base measure $Q_0$. If the SIS priors on $\bm\tau$ and $\bm\nu$ each include a complete shrinkage state that absorbs all possible chains in a finite number of steps with probability 1, then $Q, Q_1, \dotsc, Q_k \ll Q_0$ with probability 1.
\end{theorem}

\begin{remark}
A sufficient condition for the complete shrinkage state to absorb all chains in a finite number of steps with probability 1 is that the transition probability from every other state to the complete shrinkage state is bounded away from zero, which is satisfied by our choice of $\bm\Gamma$.
\end{remark}

\begin{remark}
The absorbent, complete shrinkage state in the SIS prior is needed to ensure the absolute continuity of the random distributions. \cite{opt} showed that in lieu of decreasing the Beta variances along branches of the partition tree at a sufficiently fast fixed rate \citep{Kraft1964}, one can also ensure the absolute continuity of random distributions from the P\'olya tree as long as the Beta variables will with probability 1 eventually have zero variance at deep enough levels almost everywhere on the sample space. The absorbent, complete shrinkage state in the SIS prior ensures this condition.
\end{remark}

\begin{theorem} (Prior support)
Let $f, f_1, \dotsc, f_k$ be the probability density functions corresponding to arbitrary distributions that are absolutely continuous with respect to a measure $\mu$ on $\Omega$. Let $q, q_1, \dotsc, q_k$ be corresponding densities from a HAPT model satisfying
\begin{enumerate}
\item $Q_0$ and $\mu$ are equivalent measures, that is $Q_0 \ll \mu$ and $\mu \ll Q_0$;
\item There are at least two shrinkage states, including the complete shrinkage state, at each level.
\end{enumerate}
Then $f, f_1, \dotsc, f_k$ are in the $L_1$ prior support.
\end{theorem}

Our final result gives posterior consistency in the case where the samples have equal sample sizes:

\begin{theorem} (Posterior consistency)
Let $\bm{D}_n = \{\mathbf{X}_1, \dotsc, \mathbf{X}_k\}$ be observed data consisting of $k$ independent samples, each of size $n$, from absolutely continuous distributions $P_1, \dotsc, P_k$. Let $\pi(\cdot)$ be a Hierarchical Adaptive P\'olya tree model on the $k$ densities with overall prior mean $Q_0^*$, and let $\pi(\cdot \mid \bm{D}_n)$ be the corresponding posterior. If $P_i \ll Q_0^*$ for all $i$ then we have posterior consistency under the weak topology. That is,
\[\pi(U \mid \bm{D}_n) \rightarrow 1 \text{ as } n \rightarrow \infty\]
for any weak neighborhood $U$ of the product measure $P_1 \times \dotsm \times P_k$.
\end{theorem}

\section{Methodological applications of the HAPT}
\label{methodological-applications}

We present two ways in which the HAPT model can be applied to infer structures that existing models have not been able to capture.
The first application is the ability of HAPT to model the ``dispersion function'' (defined below) on the sample space; we show how to calculate the posterior dispersion function from the $\bm\tau$ parameter.
According to our knowledge, at the time of the writing, no existing model permits inference on the variation across sample densities in this manner.
The second application is clustering samples based on their distributions, while allowing for within-cluster variation.
While the Nested Dirichlet process clusters distributions, it allows no variation among the underlying sampling distributions within each cluster. The importance of allowing for variation within clusters was first pointed out by \cite{maceachernNDP}, who described a dependent Dirichlet process which would incorporate within-cluster variation.

\subsection{Inferring the cross-sample dispersion function}
\label{variation}

The primary target of inference in problems with multiple samples is often the variation across samples.
It is this inference, for example, which lends ANOVA its name, though the model is typically presented in terms of the overall and sample means.

In the HAPT model, variation across samples is captured by $\bm\tau$, an infinite-dimensional parameter which characterizes variation at different locations and scales.
Rather than trying to provide guidance on how to interpret the multiscale structure in $\bm\tau$, we show how to recast it into an estimate of the variation across samples at any given point in the sample space, giving us a posterior \emph{dispersion function} analogous to the posterior mean function.

Let $q_\star$ be density for a new sample drawn from the HAPT model, with corresponding Beta-distributed random variables $\theta_\star(A)$ for each region $A$ in the recursive partition.
Since $q_\star$ is random, we can estimate a ``variance function'' $v: \Omega \rightarrow \mathbb{R}^+$ which gives, for any point $x$ in the sample space, the expected variance of $q_\star(x)$ conditional on the density of the common structure, $q(\cdot)$.

\begin{proposition}
The variance function is given by
\begin{align*}
&\mathbb{E}_q\left[\text{Var}\left(q_\star(x)\mid q\right)\right] = q_0(x)^2 \times \\
& \qquad \mathbb{E}_{\mathbf{S}_{\bm\tau}}\left[ \prod_{A \ni x} ||A||^2 \mathbb{E}_{\theta(A), \tau(A)}\left( \frac{\theta(A)(\theta(A)\tau(A) + 1)}{||A_\ell||^2(\tau(A) + 1)}\middle | S_\tau(A) \right)^{\mathbf{1}(x \in A_\ell)}\right. \\
&\qquad\qquad\qquad\qquad\left. \mathbb{E}_{\theta(A), \tau(A)}\left( \frac{(1-\theta(A))((1-\theta(A))\tau(A) + 1)}{||A_r||^2(\tau(A) + 1)}\middle | S_\tau(A) \right)^{\mathbf{1}(x \in A_r)} - \right.\\
& \qquad\qquad \left. \prod_{A \ni x} ||A||^2 \mathbb{E}_{\theta(A), \tau(A)}\left( \frac{\theta(A)^2}{||A_\ell||^2} \middle | S_\tau(A) \right)^{\mathbf{1}(x \in A_\ell)}\right. \\
&\qquad\qquad\qquad\qquad\left. \mathbb{E}_{\theta(A), \tau(A)}\left( \frac{(1-\theta(A))^2}{||A_r||^2} \middle | S_\tau(A) \right)^{\mathbf{1}(x \in A_r)}\right].
\end{align*}
\end{proposition}

The derivation of this result is given in the supplementary material. The expectations with respect to $\theta(A)$ and $\tau(A)$ can be estimated during the same quadrature routines used to compute the posterior distributions of $\theta(A)$, described in Section~\ref{hapt-computation}. The expectation with respect to $\bm{S_\tau}$ can then be calculated during the forward-backward routine used to calculate the posterior distribution of $\bm{S_\tau}$, as described in the same section.

We naturally expect more absolute variation across samples in areas where the densities of all the samples are higher, so we also introduce a standardized dispersion function measuring the coefficient of variation.
The posterior mean coefficient of variation of $q_\star$ at any given point is not analytically tractable; we can obtain an estimate by taking the square root of the variance function and dividing by the mean density function. We illustrate the application of the dispersion functions in Section~\ref{dispersion-process-simulations} and Section~\ref{applications}.

\subsection{Dirichlet Process Mixture of HAPT}
\label{hapt-dp-section}

Not only can the HAPT be used as a standard-alone Bayesian model, but it can also be embedded as a component in more complex hierarchical models to address a variety of inference tasks. Of course, in such cases some parts of the larger hierarchical model may require MCMC for inference, but even then, one can numerically marginalize out the HAPT portion of the model, just like one could integrate out any other conjugate model component. This often results in very simple MCMC samplers for very sophisticated nonparametric models. We provide one illustration of such an inference task in this subsection.

In many applications we may not believe that the samples collected all share a single common structure.
A more appropriate model may be that the samples are drawn from several latent populations, with samples being drawn from the same population having structure in common.
In this case we may reconstruct the latent structure by clustering the samples.
To learn the clustering of samples without fixing the number of clusters in advance, we add a Dirichlet process component to the model.
We can write the model as follows:
\begin{align*}
X_{ij} \mid Q_i &\overset{iid}{\sim} Q_i \\
Q_i \mid Q_i^*, \bm\tau_i^* &\overset{iid}{\sim} \text{PT}(Q_i^*, \bm\tau_i^*) \\
(Q_i^*, \bm\nu_i^*, \bm\tau_i^*) \mid G &\overset{iid}{\sim} G \\
G &\sim \text{DP}(\alpha H(Q^*, \bm\nu^*, \bm\tau^*)),
\end{align*}
where the base measure can be factored as
\begin{align*}
H(Q^*, \bm\nu^*, \bm\tau^*) &= \left[\pi(Q^* \mid \bm\nu^*) \times \pi(\bm\nu^*)\right] \times \pi(\bm\tau^*) \\
 &= \left[\text{PT}(Q_0, \bm\nu^*) \times \text{SIS}(\bm{\Gamma_\nu}^*)\right] \times \text{SIS}(\bm{\Gamma_\tau}^*).
\end{align*}
The Dirichlet process introduces clustering among the samples, so that some set of $Q_i$, belonging to the same cluster, have a cluster centroid $Q_i^*$ and a cluster-specific dispersion parameter, $\bm\tau_i^*$. The cluster centroids $Q_i^*$ are also allowed to have different smoothness corresponding to the cluster-specific $\bm\nu_i^*$.
Conditional on the clustering structure, the model reduces to a collection of independent HAPT models. In other words, while the above model may first appear dauntingly complex, it is nothing but a Dirichlet process mixture model on the space of distributions using the HAPT as the mixing kernel along with the hyperprior on the kernel.

We call this model a Dirichlet Process Mixture of Hierarchical Adaptive P\'olya Trees, or DPM-HAPT. 
It is comparable to the Nested Dirichlet process in the way it induces clustering among the samples, but is considerably more flexible.
While the NDP requires that all samples in a cluster have identical distributions \citep{maceachernNDP}, DPM-HAPT allows the distributions within each cluster to vary according to the HAPT model.
In addition, the advantages of the HAPT model discussed earlier, such as flexible modeling of variation in different parts of the sample space, still apply. 

Posterior computation for DPM-HAPT consists of a combination of standard Dirichlet Process methods and the HAPT posterior calculations described in Section~\ref{hapt-computation}. As noted above, conditional on the clustering structure, the model consists of a number of independent HAPT models. Although the HAPT model is not fully conjugate, our posterior computation strategy allows us to calculate the marginal likelihood to arbitrary precision. This allows the use of a Dirichlet process algorithm designed for conjugate mixture models. We use the P\'olya urn representation to sample the clustering structure, using marginal likelihoods calculated from the HAPT model.

This algorithm requires the computation of
\begin{enumerate}
\item marginal likelihoods under the HAPT model for clusters including a single sample, and
\item marginal posterior predictive likelihoods under the HAPT model of one sample conditional on one or more other samples making up a cluster.
\end{enumerate}
The first item is straightforward. The second is easily achieved by fitting the HAPT model twice, and calculating
\[f(\bm{X}_i \mid \bm{X}_{j_1}, \dotsc, \bm{X}_{j_k}) = \frac{f(\bm{X}_i, \bm{X}_{j_1}, \dotsc, \bm{X}_{j_k})}{f(\bm{X}_{j_1}, \dotsc, \bm{X}_{j_k})}.\]

\section{Simulation results}
\label{simulation}

\subsection{Density estimation and comparison to MQR}
\label{MQRcomp-section}

In this section we use simulation to evaluate the performance of density estimation under the HAPT model and compare it to the MQR model \citep{hdpm}. We would have liked to compare the performance to other models such as HDPM and NDP, but at the time of this writing there is not publicly available software for density estimation using these models. MQR is available in the {\tt R} package \texttt{DPpackage} \citep{dppackage} as the function {\tt HDPMdensity}. 

MQR models $k$ related densities with $k+1$ Dirichlet process mixtures. The density of each sample is modeled as a mixture of a common component and a unique idiosyncratic component, each with a Dirichlet process mixture prior. In all there are $k$ Dirichlet process mixtures for the $k$ idiosyncratic components and one for the common component.

\begin{figure}
\begin{center}
\includegraphics[width=\linewidth]{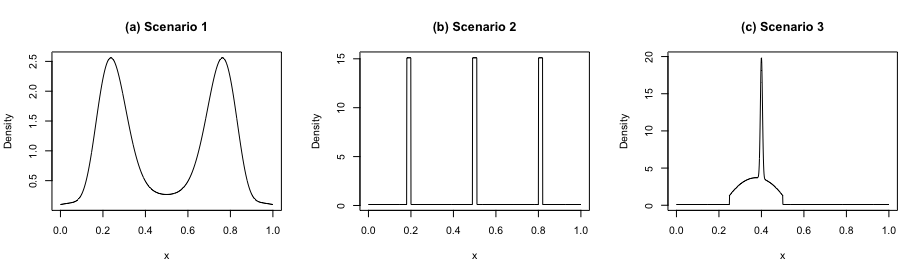}
\caption{The four-component mixtures used in the comparison between HAPT and MQR, under the expected weights. (a) shows the density for the first scenario (smooth structures easily modeled by mixtures of Normal distributions), (b) gives the density for the second scenario (narrow spikes not easily modeled by mixtures of Normal distributions), and (c) gives the density for the third scenario (both smooth structures and a narrow spike). Depending on the concentration of the Dirichlet distribution, the individual sample densities will vary more or less from this density.}
\label{MQRcomp-mixture}
\end{center}
\end{figure}

We evaluated HAPT and MQR under three simulation scenarios, each with 10 random samples of 100 observations each. In each scenario the sample densities are mixtures of four components with random weights. The first scenario constructs a smooth density out of mixtures of Beta distributions. There are two modes with weights varying from sample to sample. This density is highly amenable to modeling with the mixture of Normal distributions that MQR uses. The components are:
\begin{enumerate}
\item A Uniform(0,1) distribution, with expected weight 0.1;
\item A Beta(2,2) distribution, with expected weight 0.1;
\item A Beta(30,10) distribution, with expected weight 0.4;
\item A Beta(10,30) distribution, with expected weight 0.4.
\end{enumerate}
The density under the mean weights is illustrated in Figure~\ref{MQRcomp-mixture}(a).

The second scenario has a low-weighted diffuse base distribution, with most of the mass being concentrated in 3 spikes. We expect HAPT to perform substantially better than MQR in this scenario, as a mixture of Normal distributions does not fit the narrow spikes well. The components are:
\begin{enumerate}
\item A Uniform(0,1) distribution, with expected weight 0.1;
\item A Uniform(0.18,0.20) distribution, with expected weight 0.3;
\item A Uniform(0.49,0.51) distribution, with expected weight 0.3;
\item A Uniform(0.80,0.82) distribution, with expected weight 0.3.
\end{enumerate}
The density under the mean weights is illustrated in Figure~\ref{MQRcomp-mixture}(b)

The third scenario presents a mix of the two situations explored in the previous scenarios, with a larger-scale smooth component and a narrow spike. The components of this mixture are:
\begin{enumerate}
\item A Uniform(0,1) distribution, with expected weight 0.1;
\item A Uniform(.25,.5) distribution, with expected weight 0.3;
\item A scaled Beta(2,2) distribution, rescaled to cover the interval $[.25,.5]$, with expected weight 0.4;
\item A Beta(4000,6000) distribution, with expected weight 0.2.
\end{enumerate}
The density under the mean weights is illustrated in Figure~\ref{MQRcomp-mixture}(c)

In order to create samples with varying underlying true densities in each scenario, for each sample the weights of the mixture components were randomized according to a Dirichlet distribution with the expectations given above. The concentration of the Dirichlet distribution around the mean was varied across a wide range, with the sum of the parameters taking the values 1, 5, 10, and 50. For example, in the highest concentration setting, the weights of the four components were given by a Dirichlet(5,15,20,10) distribution.

We draw the ten underlying sample densities and then draw 100 points from each sample, for a total sample size of 1000. After fitting both models, we calculate the $L_1$ distance between each sample's true density and the estimate of the sample density given by each of the two models. This is averaged across the ten samples. We repeat this procedure twenty times for each value of the Dirichlet concentration parameter in order to gain a more accurate view of the performance of each model.

\begin{figure}
\begin{center}
\includegraphics[width=\linewidth]{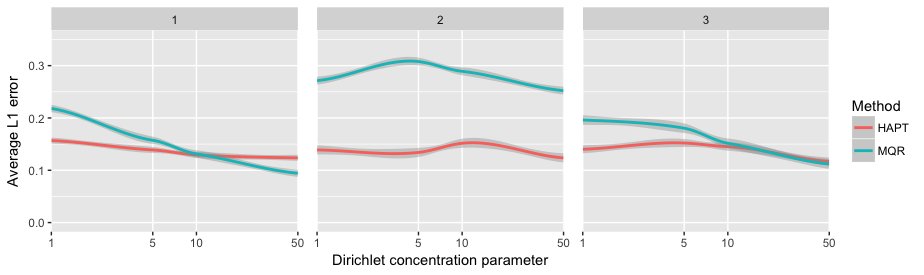}
\caption{Results of the simulation described in Section~\ref{MQRcomp-section}, comparing performance of HAPT and MQR across varying values of the Dirichlet concentration parameter. HAPT substantially outperforms MQR when the variation between samples is larger; when the sum of the Dirichlet parameters is 50 they have approximately equivalent L1 error levels, and MQR slightly outperforms HAPT when the concentration parameter is 100 or 500.}
\label{MQRcomp}
\end{center}
\end{figure}

The results of this simulation are shown in Figure~\ref{MQRcomp}. HAPT substantially outperforms MQR in all three scenarios when the variability across sample densities is large. This is likely because HAPT is still able to borrow information locally on smaller scales as appropriate, while MQR is essentially learning independent densities, with the common component having a very low weight. In the first and third scenarios MQR becomes as or more accurate than HAPT as the variability between samples decreases. This is not surprising, as MQR is able to place nearly all the weight on the common component when there is little variation across samples, resulting in almost complete shrinkage and efficient borrowing of information. The first scenario with its smooth components is particularly friendly to MQR. In the second scenario HAPT outperforms MQR by a wide margin across the board.

This simulation demonstrates that HAPT compares favorably to MQR, particularly when the densities contain features not easily modeled by a mixture of Normal distributions or the variation between samples is large. The latter observation can be attributed to HAPT's flexible nonparametric model for the variation across samples, which allows it to adjust the amount of shrinkage and information borrowing across the sample space. In contrast, MQR borrows very little information when the variation across samples is large, essentially estimating each sample's density independently. 

\subsection{Estimation of the dispersion function}
\label{dispersion-process-simulations}

\begin{figure}
\begin{center}
\includegraphics[width=\linewidth]{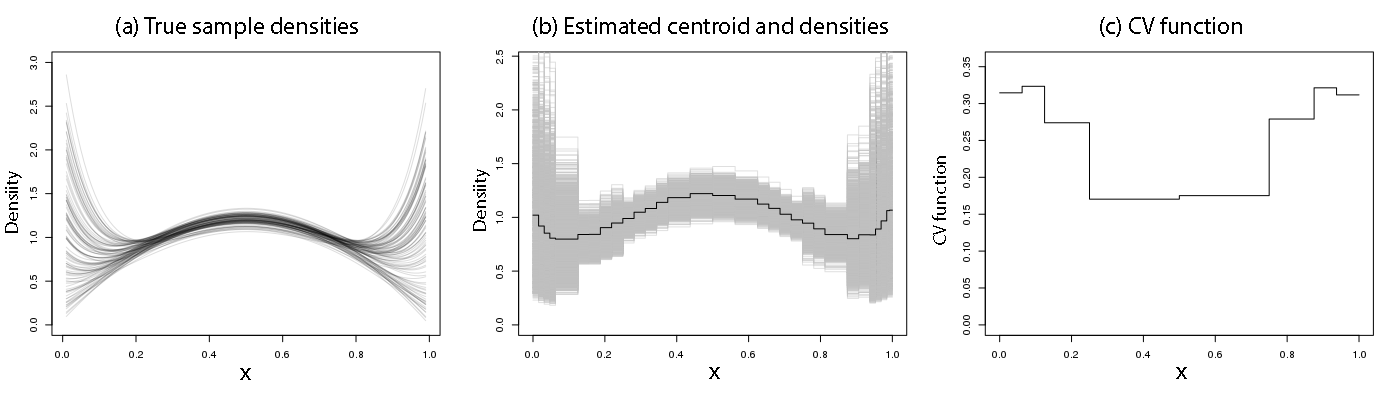}
\caption{(a) One hundred sample densities from the simulation setting used in Section~\ref{dispersion-process-simulations}; (b) Estimated centroid and sample densities after fitting the HAPT model; (c) Estimated coefficient of variation function.}
\label{het-dispersion-multi}
\end{center}
\end{figure}

To simulate data with dispersion that varies across the sample space, we simulate from a mixture of three Beta distributions, such that the variation across sample densities is low in the middle of the space and much higher near zero and one. Specifically, each sample is a mixture of three components: Beta(2,2), Beta(1,12), and Beta(12,1). The corresponding weights $w_1, w_2, w_3$ of the three components are drawn according to the following scheme. First we draw $w_1 \sim \text{Beta}(80,20)$. Then we draw $v \sim \text{Beta}(1,1)$ and set $w_2 = v(1-w_1)$ and $w_3 = (1-v)(1-w_1)$. This results in the central part of the sample space having a small amount of variation between samples, while the edges on either side have much more variation. Sample densities are illustrated in Figure~\ref{het-dispersion-multi}. One hundred sample densities are plotted in Figure~\ref{het-dispersion-multi}(a). The variation in the dispersion of the sample densities can be seen clearly.

Figure~\ref{het-dispersion-multi}(b) shows the estimated centroid and sample densities from fitting the HAPT model.
The dispersion function, or estimated coefficient of variation is plotted in Figure~\ref{het-dispersion-multi}(c).
The dispersion function clearly shows how the variation across samples is low near the center of the space and high on either end.

\subsection{DPM-HAPT Simulations}
\label{hapt-dp-simulations}

\begin{figure}
\begin{center}
\includegraphics[width=\linewidth]{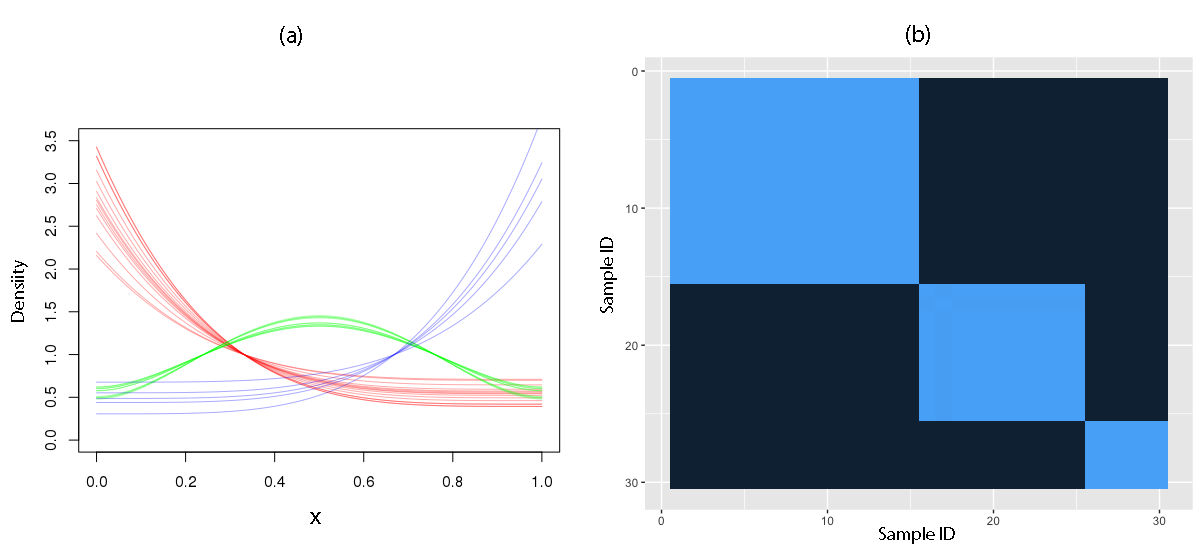}
\caption{(a) Sample densities for the clustering simulation, and (b) probability of two samples clustering together based on the DPM-HAPT model, which clearly identifies the three clusters that exist in this simulation, despite significant variation across samples within clusters.}
\label{coclustering}
\end{center}
\end{figure}

We simulate a simple 1-dimensional example to demonstrate the clustering behavior of the DPM-HAPT model. The simulation contains 30 samples belonging to three true clusters, with 15, 10, and 5 samples respectively. Each sample is drawn from a mixture of a Uniform(0,1) distribution and a Beta distribution, with the parameters of the Beta varying by cluster: Beta(1,5) for the first cluster, Beta(3,3) for the second cluster, and Beta(5,1) for the third cluster. The weights of the two components are randomized in each sample. The weight of one component is drawn from a Beta(10,10) distribution, which creates weights varying approximately between 0.3 and 0.7, with the actual observed proportions in realized samples varying more widely due to the additional Binomial variation. Sample densities are plotted in Figure~\ref{coclustering}(a). Each sample contains $n=300$ points.

An MCMC sampler is run using the P\'olya urn scheme to sample the clustering structure. We summarize the results by looking at how often each of the 30 samples is clustered together with each other sample. These results are plotted in Figure~\ref{coclustering}(b). We can see in the figure that the DPM-HAPT model clearly identified the three clusters.

\begin{figure}
\begin{center}
\includegraphics[width=\linewidth]{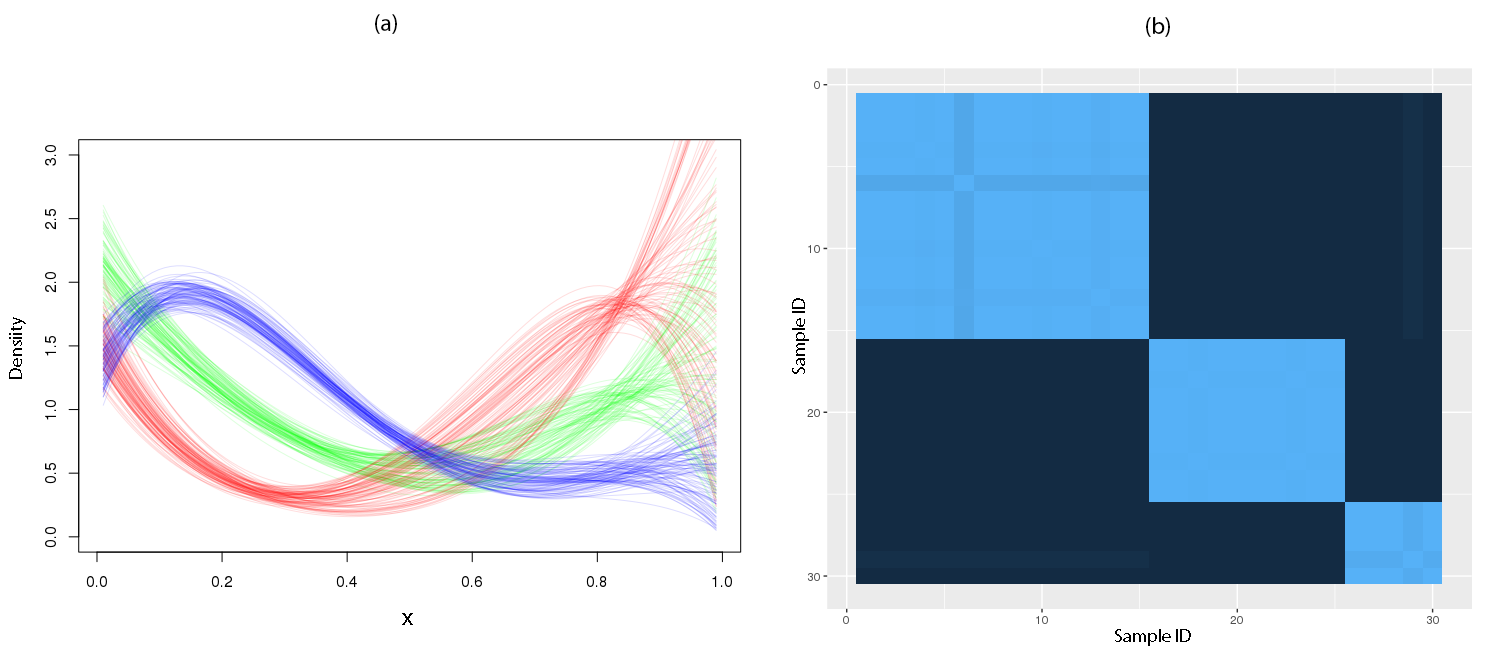}
\caption{(a) One hundred draws from each of three clusters in the heterogeneous variation clustering example; (b) Probability of two samples clustering together based on 1,000 MCMC draws for the heterogeneous variance example. Despite the variation in the dispersion of the densities, HAPT-DPM clearly identifies the three true clusters.}
\label{het-cluster-multi}
\end{center}
\end{figure}

We now consider an example in which the variation across samples differs across the sample space.
Samples with heterogeneous variation are drawn from mixtures of four Beta distributions, with varying parameters for each cluster. 
The parameter are chosen so that each cluster has much higher variation across sample densities on the right half of the space than on the left half.
The data generating process is described in the supplementary material.
We consider 30 samples belonging to three true clusters, as above. Figure~\ref{het-cluster-multi} (a) shows one hundred draws from each of three clusters used in this simulation, to illustrate the variability between and within clusters and the heterogeneity across the sample space.

As in the previous example, we simulate three clusters with 15, 10 and 5 samples respectively. We draw 150 observations from each sample, and apply DPM-HAPT to cluster the resulting samples.
We run the MCMC for 1,000 draws after burnin; estimated coclustering probabilities are plotted in Figure~\ref{het-cluster-multi} (b).
The three true clusters are clearly identified even in the presence of substantial heterogeneity.

\section{Application: DNase-seq data}
\label{applications}
DNase sequencing (DNase-seq) is a method used to identify regulatory regions of the genome \citep{Song2010}. DNA is treated with Deoxyribonuclease~(DNase)~I, an enzyme that cuts the DNA strand. The cut strands are then sequenced and the locations of the cuts are identified and tallied. The vulnerability of the DNA strand to DNase varies by location, resulting in a distribution of cut counts which is nonuniform. The density of this distribution is related to various biological factors of interest: for example, it tends to be high near potential binding sites for transcription factors, since these proteins require access to the DNA strand in much the same way as DNase~I, but will be low if a transcription factor already bound at that site blocks access to the DNA strand. 

We consider the problem of clustering DNase-seq profiles near potential transcription binding sites, identified by a specific genetic motif. Each sample consists of observed counts in a range of 100 base pairs on either side of one occurrence of the motif. A single motif, consisting of 10--20 base pairs, may appear tens of thousands of time in the genome, with each occurrence presenting one sample for analysis. Many samples, however, have very few cuts observed. For analysis we restrict ourselves to samples which meet a minimum sample size threshold.

Different locations where the transcription factor motif of interest appears may be expected to show different DNase behavior in the region around the motif for a variety of reasons. This makes clustering a more appropriate approach to the problem than treating all the samples as having a single common structure. Identifying clusters of locations which have similar DNase-seq profiles may reveal previously unrecognized factors. We also expect within-cluster variation above and beyond sampling variation, which makes the Nested Dirichlet process unsuitable.

Here we present data from the ENCODE project \citep{ENCODEProjectConsortium2012} for locations surrounding a motif associated with the RE1-silencing transcription factor (REST). REST suppresses neuronal genes in non-neuronal cells \citep{Chong1995}. The data includes 48,549 locations where the REST motif appears in the genome. The motif consists of 21 base pairs, and the data includes an additional 100 base pairs on each side, for a total of 221 base pairs. In all, 922,704 cuts were tallied, an average of 19 per location. 468 locations have zero cuts observed. The number of cuts per location is heavily right skewed, with a median of 13 observations, first and third quartiles of 7 and 21 respectively, and a maximum of 2,099 cuts observed in a single sample.

For this analysis we restrict ourselves to locations which have at least 200 observations, a total of 265 samples. These samples include a total of 70,225 observations, an average of 330 observations per sample. The distribution is still quite skewed, with a minimum of 201 observations and a maximum of 2099. The median is 279 and the first and third quartiles are 232 and 366 observations. Histograms of 30 of the 265 locations are shown in the supplementary material.

\begin{figure}
\begin{center}
\includegraphics[width=\linewidth]{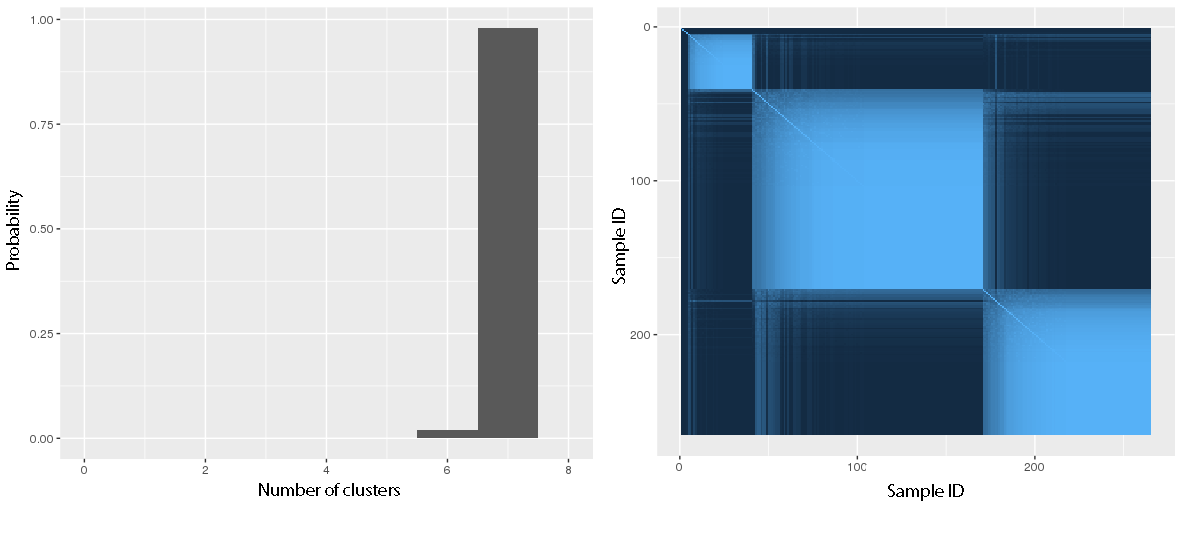}
\caption{(a) Posterior distribution of the number of clusters from the DPM-HAPT in the DNase-seq application; (b) the model shows clear clustering of the samples.}
\label{REST-cluster-hist}
\end{center}
\end{figure}

We fit the DPM-HAPT clustering model to this data, using 100 post-burnin draws for inference. The model estimates 7 clusters with high probability (see Figure~\ref{REST-cluster-hist}(a)), of which there are three large clusters and four singleton locations, each of which consists of a single large spike. One of the singleton locations occasionally joins one of the larger clusters, resulting in 6 clusters.
A heatmap of the clustering structure is shown in Figure~\ref{REST-cluster-hist}(b).
The clusters are clearly differentiated and vary in size, with the largest cluster containing about 130 locations, though the cluster sizes vary from iteration to iteration due to uncertainty in the cluster assignment.

\begin{figure}[htb]
\begin{center}
\includegraphics[width=\linewidth]{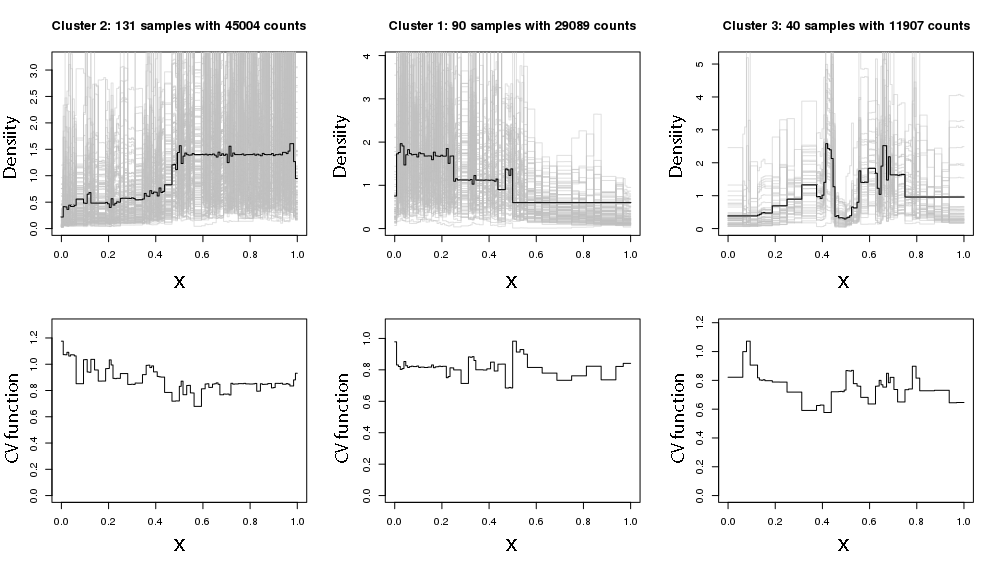}
\caption{Posterior mean densities and estimated dispersion functions of the three largest clusters in the DNase-seq example. The heavy black line shows the cluster centroid; light gray lines in the background show the estimated means of each sample in the cluster. The dispersion function is plotted below.}
\label{REST-cluster-densities}
\end{center}
\end{figure}

The estimated posterior mean densities of the three largest clusters are plotted in Figure~\ref{REST-cluster-densities}.
One of these clusters includes locations with cuts which are roughly symmetric around the transcription factor binding site with a dip in the middle. The central dip indicates that a transcription factor may be bound at that location, blocking access for the DNase~I molecule. The other two largest clusters include locations which have cuts heaped up on one side or another of the binding site, suggestive of larger-scale patterns in the DNase~I sensitivity due to factors like the folded structure of the DNA. The four singleton clusters show other densities which do not conform to the general patterns of the three largest clusters. The plots show substantial variation around the cluster centroid, much more than can be explained by sampling variation alone. The corresponding estimated dispersion functions are plotted in the second row, and are quite noisy; this is not surprising given that they are cross-sample dispersions (a second-order parameter) estimated with relatively few degrees of freedom.

\section{Discussion}
\label{discussion}

The HAPT model offers a compelling alternative to existing nonparametric models that share information across multiple samples.
The P\'olya tree's ability to directly model the density of an absolutely continuous distribution frees us from the necessity of using mixture models to obtain densities, while the computational tractability of the posterior avoids the need to run MCMC chains for posterior inference.
The addition of the SIS prior allows us not only to more accurately model the densities of interest, but also to estimate a fully nonparametric dispersion function over the sample space.
The model extends easily---both conceptually and computationally---to the setting where we do not believe all our samples have the same common structure, where DPM-HAPT allows us to learn both clustering structure and the distributional structure within each cluster.

Although we have presented HAPT in a one-dimensional space for the sake of clarity, adoption of the randomized recursive partitioning scheme first introduced in \cite{opt} allows the extension of HAPT to model densities in multidimensional spaces.
Variables other than simple continuous ones can also be handled naturally---all that is needed is the definition of an appropriate recursive partition.
This allows inclusion of categorical and ordinal-valued variables, as well has more exotic possibilities: a continuous variable that lives on the surface of a torus, a partially ordered categorical variable, or a zero-inflated variable with a point mass at zero and a continuous component on the positive halfline.

The P\'olya tree's decomposition of the density space into orthogonal Beta-distributed random variables, which extends to HAPT, is central to HAPT's computational efficiency.
It also allows the performance of quick online updates in the HAPT model: when a new data point arrives we only need to update the nodes of the P\'olya tree which contain the new data point, rather than recomputing the entire posterior.
In a HAPT truncated at a depth of $L$ levels, this means we need to reevaluate the posteriors of only $L$ nodes, rather than $2^L$.
HAPT may thus be used in streaming data settings where fast online updates are essential. This is one of our current developments.

The DPM-HAPT model illustrates the flexibility of HAPT as a component of larger hierarchical models as well as the computational tractability resulting from the ``almost'' conjugacy of the HAPT. Another such example is a mixture of HAPT model. Specifically, one well-recognized limitation of the P\'olya tree and related models is its dependence on the choice of a particular partition tree on the sample space, which is typically determined by the quantiles of a given base measure $Q_0$. \cite{Hanson2006} shows that by treating the base measure, and thus the partition tree, as an unknown and place a hyperprior on it, one can arrive at the so-called mixture of P\'olya trees, which overcomes the sensitivity to the choice of the partition, and moreover, very simple MCMC strategies can be used for such models because the P\'olya tree can be analytically marginalized out given each base measure. We note that exactly the same extension can be applied to the HAPT by placing hyperpriors on the base measure. The same samplers that work for the mixture of P\'olya trees will work for the mixture of HAPT as the HAPT also be integrated out numerically. 

Finally, we note that the DPM-HAPT model may also be easily extended by replacing the Dirichlet process with any cluster-inducing process which admits inference given the marginal likelihoods and conditional probabilities of the clusters.
This includes, for example, the Pitman-Yor process.
This allows the properties of the clustering process to be adapted if the clustering assumptions implicit in the Dirichlet process are not appropriate.	

\section*{Software}
An {\tt R} package, {\tt HAPT-package}, which implements the HAPT model is available at \url{https://github.com/MaStatLab/HAPT-package}.

\section*{Acknowledgment}
LM's research is supported by NSF grants DMS-1612889 and DMS-1749789. 

\bibliographystyle{rss}
\bibliography{hapt}

\newpage

\setcounter{page}{1}

\beginsupplement
\section*{Supplementary material}

\subsection*{S1: Posterior derivation}
\label{hapt-posterior}

We use the following notation in this section:
\begin{itemize}
\item $A$ is a set in the recursive partition of $\Omega$ implicit in the P\'olya tree.
\item $A_\ell$ and $A_r$ indicate the left and right children of $A$, respectively; these are also sets in the recursive partition of $\Omega$.
\item $n(A)$ is the number of data points across all samples that are contained in $A$; $n_i(A)$ is the number of datapoints in sample $i$ that are contained in $A$. Thus $n(A) = \sum_i n_i(A)$.
\end{itemize}
We ignore for the time being the priors on $\bm\tau$ and $\bm\nu$, as they are not important for this derivation and can be reinserted at the end.

The density of the distribution generated from the HAPT model at an observation $x_{ij}$ consists of three factors:
\begin{enumerate}
\item The baseline density $q_0(x_{ij})$
\item A term indicating how the common structure modifies the density: 
\[\prod_{A: x_{ij}\in A_\ell} \theta(A) \frac{Q_0(A)}{Q_0(A_\ell)} \prod_{A: x_{ij}\in A_r} (1-\theta(A)) \frac{Q_0(A)}{Q_0(A_r)}\]
Note that under the canonical representation, $\frac{Q_0(A)}{Q_0(A_\ell)} \equiv 2$
\item A term indicating how the idiosyncratic structure of the particular sample containing $x$ modifies the density:
\[\prod_{A: x_{ij}\in A_\ell} \theta_i(A) \frac{Q(A)}{Q(A_\ell)} \prod_{A: x_{ij}\in A_r} (1-\theta_i(A)) \frac{Q(A)}{Q(A_r)}\]

Note that by definition, $\frac{Q(A)}{Q(A_\ell)} = 1/\theta(A)$ and $\frac{Q(A)}{Q(A_r)} = 1/(1-\theta)$.
\end{enumerate} 
Altogether this gives us the following likelihood:
\begin{align*}
f(\mathbf{X} \mid \mathbf{\theta}, \mathbf{\theta}_i) &= \prod_{x_{ij}} \left[ q_0(x_{ij}) \prod_{A: x_{ij}\in A_\ell} \left[ \theta(A) \frac{Q_0(A)}{Q_0(A_\ell)} \right] \prod_{A: x_{ij}\in A_r} \left[ (1-\theta(A)) \frac{Q_0(A)}{Q_0(A_r)} \right]\right. \\
&\qquad\qquad\qquad\qquad \left.\prod_{A: x_{ij}\in A_\ell} \left[ \theta_i(A) \frac{Q(A)}{Q(A_\ell)} \right] \prod_{A: x_{ij}\in A_r} \left[ (1-\theta_i(A)) \frac{Q(A)}{Q(A_r)} \right] \right] \\
&= \prod_{x_{ij}} \left[ q_0(x_{ij}) \prod_{A: x_{ij}\in A_\ell} \left[ \theta_i(A) \frac{Q_0(A)}{Q_0(A_\ell)} \right] \prod_{A: x_{ij}\in A_r} \left[ (1-\theta_i(A)) \frac{Q_0(A)}{Q_0(A_r)} \right] \right]. \\
\end{align*}

Rearranging terms, the likelihood can be written as follows:
\begin{align*}
&f(\mathbf{X}\mid \theta_i, \theta) = \prod_{x_{ij}} q_0(x_{ij}) \times \\
&\qquad\prod_A \left[ \prod_{i=1}^k \left[\theta_i(A)\frac{Q_0(A)}{Q_0(A_\ell)}\right] ^{n_i(A_\ell)} \left[(1-\theta_i(A))\frac{Q_0(A)}{Q_0(A_r)}\right] ^{n_i(A_r)}  \right],
\end{align*}
which gives the following form for the posterior of $\theta(A), \theta_i(A)$ for a particular $A$:
\begin{align*}
\pi(\theta(A),\theta_i(A) \mid \tau(A), \nu(A), x) &\propto \theta(A) ^{\theta_0(A)\nu(A) -1} (1-\theta(A)) ^{(1-\theta_0(A))\nu(A)-1}\times\\ &\left[\frac{\Gamma(\tau(A))}{\Gamma(\theta(A)\tau(A))\Gamma((1-\theta(A))\tau(A))}\right]^k \times \\
&\prod_{i=1}^k \left[ \theta_i(A) ^{\theta(A)\tau(A) + n_i(A_\ell) -1} (1-\theta_i(A)) ^{(1-\theta(A))\tau(A) + n_i(A_r) -1}\right].
\end{align*}
Conditional on $\theta(A)$ and $\tau(A)$ the $\theta_i(A)$ are Beta distributed, and we can integrate them out analytically:
\begin{align*}
\pi(\theta(A) \mid \tau(A), \nu(A), x) &\propto \theta(A) ^{\theta_0(A)\nu(A)-1} (1-\theta(A)) ^{(1-\theta_0(A))\nu(A) -1} \times \\ 
&\qquad \left[\frac{\Gamma(\tau(A))} {\Gamma(\theta(A)\tau(A))\Gamma((1-\theta(A))\tau(A))}\right]^k \times \\
&\qquad \prod_{i=1}^k \frac{\Gamma\left(\theta(A) \tau(A) + n_i (A_\ell)\right) \Gamma\left((1-\theta(A)) \tau(A) + n_i (A_r)\right)}{\Gamma\left(\tau(A) + n_i(A)\right)}.
\end{align*}
We can then simply multiply by the priors on $\tau(A)$ and $\nu(A)$ to obtain
\begin{align*}
\pi(\theta(A), \tau(A), \nu(A) &\mid S_\tau(A), S_\nu(A), x) \propto \theta(A) ^{\theta_0(A)\nu(A) -1} (1-\theta(A)) ^{(1-\theta_0(A))\nu(A) -1}\times \\
&\qquad \left[\frac{\Gamma(\tau(A))}{\Gamma(\theta(A)\tau(A))\Gamma((1-\theta(A))\tau(A))}\right]^k \times \\
&\qquad \prod_{i=1}^k \frac{\Gamma\left(\theta(A) \tau(A) + n_i (A_\ell)\right) \Gamma\left((1-\theta(A)) \tau(A) + n_i (A_r)\right)}{\Gamma\left(\tau(A) + n_i(A)\right)} \times\\
&\qquad \pi(\tau(A))\pi(\nu(A)).
\end{align*}

\subsection*{S2: Proofs of theoretical results}
\label{hapt-proofs}

\textsc{Proof of Theorem 1: Absolute continuity.} The result follows directly from repeated application of Theorem 3 in \cite{apt}. From one application of that theorem we have that $Q \ll Q_0$ with probability 1; by a second application we have that $Q_i \ll Q$ for each $Q_i$ with probability 1. Thus $Q_i \ll Q_0$ with probability 1.

\textsc{Proof of Theorem 2: Large L1 Support.} The upper level of the hierarchy ($q$) follows immediately from Theorem 4 in \cite{apt}, since its prior coverage is not altered by the addition of the hierarchy. For the second level of the hierarchy, note that by Theorem 4 in \cite{apt} we have
\[\forall q \ll \mu,\; P(\int |q_i(x) - f_i(x)| d\mu < \tau_i\mid q) > 0.\]
Further, by definition of conditional expectation, 
\[P\left(\int |q_i(x) - f_i(x)| d\mu < \tau_i\mid q\right) = E\left(\mathbf{1}\left[\int |q_i(x) - f_i(x)| d\mu < \tau_i\right]\mid q\right)\]
is measurable with respect to $Q$.  Thus
\[
P\left(\int |q_i(x) - f_i(x)| d\mu < \tau_i\right) = \int P\left(\int |q_i(x) - f_i(x)| d\mu < \tau_i\mid q\right) \; dQ > 0,
\]
as the integrand is positive with probability 1. 

\textsc{Proof of Theorem 3: Weak Consistancy.} By Schwartz's theorem, it is sufficient to show that $P_1, \dotsc, P_k$ jointly lie in the Kullback-Leibler support of the prior. The proof proceeds as follows: We consider the product space of all $k$ distributions so that we can show joint convergence. We restrict ourselves to a compact set with mass $1-\epsilon'$, and define a set $\tilde D_\epsilon$, which depends on $p_0$. We show that $\tilde D_\epsilon$ has positive prior mass. We then show that by choosing $\epsilon$ and $\epsilon'$ appropriately, we can make $\tilde D_\epsilon$ lie within an arbitrarily small K-L ball around $p_0$. This shows that $P_1, \dotsc, P_k$ are jointly in the support of the prior, and concludes the proof. We follow closely the proof of Theorem 5 in \cite{apt}.

Let $P_0$ be the product measure $P_0 = P_1 \otimes \dotsm \otimes P_k$ on $\Omega^k$, and let $p_0 = dP_0/d\mu$ be the corresponding density. Let $Q_0 = Q_0^* \otimes \dotsm \otimes Q_0^*$, and let $q_0 = dQ_0/d\mu$. Define additionally the densities $\tilde p_0 = dP_0/dQ_0$ and $\tilde q = dQ/dQ_0$ for any $Q \ll Q_0$. Let $M$ be a finite upper bound on $\tilde p_0$. The Kullback-Leibler distance between $p_0$ and $q$ is given by
\[\text{KL}_\mu(p_0,q) = \int p_0 \log (p_0/q)\, d\mu = \int \tilde p_0 \log(\tilde p_0/\tilde q)\, dQ_0 = \text{KL}_{Q_0}(\tilde p_0, \tilde q).\]
By Lusin's theorem, for any $\epsilon' > 0$ there exists a compact set $E \subset \Omega$ with $Q_0(E^c) < \epsilon'$, such that $\tilde p_0$ is continuous (and so uniformly continuous) on $E$. For any $\epsilon > 0$ there exists a partition $\Omega = \cup_i A_i$ with all $A_i \in \mathcal{A}^(k)$ for some $k$, such that the diameter of each $A_i\cap E$ is less than $\epsilon$. We define
\[\delta_E(\epsilon) = \sup_{x,y \in E:|x-y|<\epsilon} |\tilde p_0(x) - \tilde p_0(y)|\]
and
\[d_i = \max \left( \sup_{A_i \cap E} \tilde p_0(x) + \delta_E(\epsilon), \epsilon'\right).\]
Note that because $p_0$ is uniformly continuous on $E$, $\delta_E(\epsilon) \rightarrow 0$ as $\epsilon \rightarrow 0$.

Let $D_\epsilon(\tilde p_0)$ be the collection of step functions $g(x) = \sum_i g_i \textbf{1}_{A_i}(x)$ with $d_i \leq g_i < d_i + \delta_E(\epsilon)$. For every $g \in D_\epsilon(\tilde p_0)$, let $\tilde g = g/\int g\, dQ_0$ be the normalized version of $g$, and let $\tilde D_\epsilon(\tilde p_0)$ be the collection of the $\tilde g$.

Let $I$ be the number of sets $A_i$ in the partition $\Omega = \cup_i A_i$. We can consider each step function $g$ as a point in $I$-dimensional space, where the $i$th dimension corresponds to the value of the step function on the set $A_i$. Note that $D_\epsilon(\tilde p_0) = [d_1, d_1 + \delta_E(\epsilon)) \times \dotsm \times [d_I, d_I + \delta_E(\epsilon))$ is a convex set in this $I$-dimensional space, and we can find an open ball in this $I$-dimensional space which is a subset of $D_\epsilon(\tilde p_0)$. The normalized version $\tilde D_\epsilon(\tilde p_0)$ is in turn a convex set in the $(I-1)$-simplex, which also contains an open ball in the $(I-1)$-simplex.

The HAPT model places positive probability on normalized step functions taking unique values precisely on the sets $A_i$, i.e. the same $(I-1)$-simplex noted above. Because each of the mass assignment parameters in HAPT has a Beta prior, HAPT has positive prior density everywhere in the $(I-1)$ simplex. Because $\tilde D_\epsilon(\tilde p_0)$ contains an open ball, it follows that it has positive prior mass.

It now remains to show that we can bound $\tilde D_\epsilon(\tilde p_0)$ within an arbitrarily small ball about $p_0$. The remainder of the proof follows exactly as Ma (2016). For every $\tilde g \in \tilde D_\epsilon(\tilde p_0)$, we have
\begin{align*}
0 \leq \text{KL}_{Q_0}(\tilde p_0, \tilde g) &= \int \tilde p_0 \log (\tilde p_0/ \tilde g) \, dQ_0 \\
  &= \int \tilde p_0 \log (\tilde p_0/ g) \, dQ_0 + \log \left( \int g\, dQ_0\right) \\
  &= \int_E \tilde p_0 \log (\tilde p_0/ g) \, dQ_0 + \int_{E^c} \tilde p_0 \log (\tilde p_0/ g) \, dQ_0 + \log \left( \int g\, dQ_0\right).
\end{align*}
Because $g \geq p_0$ everywhere in $E$, the first integral is not greater than 0. The second integral is bounded by $M \log (M / \epsilon') \epsilon'$, which goes to zero as $\epsilon' \rightarrow 0$. To bound the third term, we note that
\[ \log \left( \int g\, dQ_0\right) = \log \left( 1 + \int (g - \tilde p_0)\, dQ_0\right) \leq \int (g - \tilde p_0)\, dQ_0,\]
We can bound this last integral by
\[\int (g - \tilde p_0)\, dQ_0 \leq \int_E (g - \tilde p_0)\, dQ_0 + \int_{E^c} |g - \tilde p_0|\, dQ_0.\]
On the set $E$, we have $g_0 - \tilde p_0 \leq 3\delta_E(\epsilon) + \epsilon'$, and on $E^c$ we have $g_0 - \tilde p_0 \leq 3M + \epsilon'$. Thus
\[\int (g - \tilde p_0)\, dQ_0 \leq 3\delta_E(\epsilon) + \epsilon' + (3M + \epsilon')\epsilon' \rightarrow 0 \text{ as } \epsilon, \epsilon' \rightarrow 0.\]
This shows that by picking $\epsilon, \epsilon'$ appropriately, $\tilde D_\epsilon(\tilde p_0)$ is contained within an arbitrarily small KL ball about $\tilde p_0$, and so $p_0$ is in the KL support of the HAPT model.

\subsection*{S3: Derivation of the variance function}

We have
\[q_\star(x) = q_0(x) \prod_{A \ni x} ||A|| \left(\frac{\theta_\star(A)}{||A_\ell||}\right)^{\mathbf{1}(x \in A_\ell)} \left(\frac{1-\theta_\star(A)}{||A_r||}\right)^{\mathbf{1}(x \in A_r)}, \]
where $q_0(\cdot)$ is the density corresponding to the prior mean distribution $Q_0$. We note that 
\[\theta_\star(A) | \theta(A), \tau(A) \sim \text{Beta}(\theta(A)\tau(A), (1-\theta(A))\tau(A)).\]
We will need the facts that 
\[\mathbb{E}(\theta_\star(A) \mid \theta(A), \tau(A)) = \theta(A)\]
and
\[\text{Var}(\theta_\star(A) \mid \theta(A), \tau(A)) = \text{Var}(1-\theta_\star(A) \mid \theta(A), \tau(A) ) = \frac{\theta(A)(1-\theta(A))}{\tau(A)+1},\]
which together imply
\begin{align*}
\mathbb{E}(\theta_\star(A)^2 \mid \theta(A), \tau(A)) &= \frac{\theta(A)(1-\theta(A))}{\tau(A)+1} + \theta(A)^2 \\
 &= \frac{\theta(A)(\theta(A)\tau(A) + 1)}{\tau(A) + 1}\\
\mathbb{E}((1-\theta_\star(A))^2 \mid \theta(A), \tau(A)) &= \frac{\theta(A)(1-\theta(A))}{\tau(A)+1} + (1-\theta(A))^2 \\
 &= \frac{(1-\theta(A))((1-\theta(A))\tau(A) + 1)}{\tau(A) + 1}.
\end{align*}

Let $\bm{\theta}(x) = \left\{\theta(A): A \ni x\right\}$. It is clear that $\bm{\theta}(x)$ contains exactly the information about $q$ which is relevant to $q_\star(x)$; that is, the distribution of $q_\star(x) \mid \bm\theta(x)$ is identical to the distribution of $q_\star(x) \mid q$. With the above, we can calculate
\begin{align*}
&\text{Var}\left(q_\star(x) \mid q\right) = \mathbb{E}_{\mathbf{S}_{\bm\tau}, \bm\tau} \left[\text{Var}\left(q_\star(x)\mid q, \bm\tau, \mathbf{S}_{\bm\tau}\right)\mid q\right] + \text{Var}_{\mathbf{S}_{\bm\tau}, \bm\tau} \left[\mathbb{E}\left(q_\star(x)\mid q, \bm\tau, \mathbf{s}_{\bm\tau}\right)\mid q\right] = \\
  &\quad \!\! q_0(x)^2 \mathbb{E}_{\mathbf{S}_{\bm\tau}, \bm\tau}\left[ \text{Var} \left( \prod_{A \ni x} ||A|| \left(\frac{\theta_\star(A)}{||A_\ell||}\right)^{\mathbf{1}(x \in A_\ell)} \left(\frac{1-\theta_\star(A)}{||A_r||}\right)^{\mathbf{1}(x \in A_r)} \middle| \bm{\theta}(x), \bm\tau, \mathbf{S}_{\bm\tau}\right) \middle| \bm{\theta}(x)\right] + \\
  &\quad \!\! q_0(x)^2 \text{Var}_{\mathbf{S}_{\bm\tau}, \bm\tau}\left[\mathbb{E} \left(\prod_{A \ni x} ||A|| \left(\frac{\theta_\star(A)}{||A_\ell||}\right)^{\mathbf{1}(x \in A_\ell)} \left(\frac{1-\theta_\star(A)}{||A_r||}\right)^{\mathbf{1}(x \in A_r)} \middle| \bm{\theta}(x), \bm\tau, \mathbf{S}_{\bm\tau}\right) \middle| \bm{\theta}(x)\right].
\end{align*}
We note that the expectation in the second term can be factored into a product of expectations, none of which depend on $\bm\tau$ or $\bm{S_\tau}$.
The variance of this product is thus zero.
We rewrite the remaining line using the identity $\text{Var}(X) = \mathbb{E}(X^2) - \mathbb{E}(X)^2$:
\begin{align*}
&\text{Var}\left(q_\star(x)\mid q\right) = q_0(x)^2 \times \\
&\qquad\mathbb{E}_{\mathbf{S}_{\bm\tau}, \bm\tau}\left[ \mathbb{E} \left( \left( \prod_{A \ni x} ||A|| \left(\frac{\theta_\star(A)}{||A_\ell||}\right)^{\mathbf{1}(x \in A_\ell)} \left(\frac{1-\theta_\star(A)}{||A_r||}\right)^{\mathbf{1}(x \in A_r)}\right)^2 \middle| \bm{\theta}(x), \bm\tau, \mathbf{S}_{\bm\tau}\right) - \right.\\
&\qquad \left. \left( \mathbb{E} \left( \prod_{A \ni x} ||A|| \left(\frac{\theta_\star(A)}{||A_\ell||}\right)^{\mathbf{1}(x \in A_\ell)} \left(\frac{1-\theta_\star(A)}{||A_r||}\right)^{\mathbf{1}(x \in A_r)} \middle| \bm{\theta}(x), \bm\tau, \mathbf{S}_{\bm\tau}\right)\right)^2 \middle| \bm{\theta}(x)\right]. \\
\end{align*}
 Conditional on $\bm\theta(x)$ and $\bm{S_\tau}$, the terms in both products are mutually independent, both \emph{a priori} and \emph{a posteriori}.
Thus, we can factor the expectations:
\begin{align*}
&\text{Var}\left(q_\star(x)\mid q\right) = q_0(x)^2 \times \\
&\quad\mathbb{E}_{\mathbf{S}_{\bm\tau}, \bm\tau}\left[ \prod_{A \ni x} \mathbb{E} \left( \left( ||A|| \left(\frac{\theta_\star(A)}{||A_\ell||}\right)^{\mathbf{1}(x \in A_\ell)} \left(\frac{1-\theta_\star(A)}{||A_r||}\right)^{\mathbf{1}(x \in A_r)}\right)^2 \middle| \theta(A), \tau(A), S_\tau(A)\right) - \right.\\
& \quad \left. \prod_{A \ni x} \left( \mathbb{E} \left( ||A|| \left(\frac{\theta_\star(A)}{||A_\ell||}\right)^{\mathbf{1}(x \in A_\ell)} \left(\frac{1-\theta_\star(A)}{||A_r||}\right)^{\mathbf{1}(x \in A_r)} \middle| \theta(A), \tau(A), S_\tau(A)\right)\right)^2 \middle| \bm{\theta}(x)\right].
\end{align*}
We can now plug in our expressions for the expectations, calculated above:
\begin{align*}
&\text{Var}\left(q_\star(x)\mid q\right) = q_0(x)^2 \times \\ 
&\qquad \mathbb{E}_{\mathbf{S}_{\bm\tau}, \bm\tau}\left[ \prod_{A \ni x} ||A||^2 \left( \frac{\theta(A)(\theta(A)\tau(A) + 1)}{||A_\ell||^2(\tau(A) + 1)} \right)^{\mathbf{1}(x \in A_\ell)}\right. \\
&\qquad\qquad\qquad\qquad\quad\left.\left( \frac{(1-\theta(A))((1-\theta(A))\tau(A) + 1)}{||A_r||^2(\tau(A) + 1)} \right)^{\mathbf{1}(x \in A_r)} - \right.\\
& \qquad\qquad\quad \left. \prod_{A \ni x} ||A||^2 \left( \frac{\theta(A)^2}{||A_\ell||^2} \right)^{\mathbf{1}(x \in A_\ell)} \left( \frac{(1-\theta(A))^2}{||A_r||^2} \right)^{\mathbf{1}(x \in A_r)} \middle| \bm{\theta}(x)\right].
\end{align*}
Since $q$ is itself unknown, we take the expectation with respect to it:
\begin{align*}
&\mathbb{E}_q\left[\text{Var}\left(q_\star(x)\mid q\right)\right] = q_0(x)^2 \times \\
&\qquad\mathbb{E}_{\mathbf{S}_{\bm\tau}, \bm\tau, \bm\theta(x)}\left[ \prod_{A \ni x} ||A||^2 \left( \frac{\theta(A)(\theta(A)\tau(A) + 1)}{||A_\ell||^2(\tau(A) + 1)} \right)^{\mathbf{1}(x \in A_\ell)}\right. \\
&\qquad\qquad\qquad\qquad\qquad\quad\left. \left( \frac{(1-\theta(A))((1-\theta(A))\tau(A) + 1)}{||A_r||^2(\tau(A) + 1)} \right)^{\mathbf{1}(x \in A_r)} - \right.\\
& \qquad\qquad\qquad\quad \left. \prod_{A \ni x} ||A||^2 \left( \frac{\theta(A)^2}{||A_\ell||^2} \right)^{\mathbf{1}(x \in A_\ell)} \left( \frac{(1-\theta(A))^2}{||A_r||^2} \right)^{\mathbf{1}(x \in A_r)}\right].
\end{align*}
We call this expectation the {\it variance function}. Finally, we note that conditional on $\mathbf{S}_{\bm\tau}$, $\theta(A), \tau(A)$ are independent of $\theta(A'), \tau(A')$ for any two distinct regions $A$ and $A'$. This allows us to rearrange our expectation to obtain
\begin{align*}
&\mathbb{E}_q\left[\text{Var}\left(q_\star(x)\mid q\right)\right] = q_0(x)^2 \times \\
& \qquad \mathbb{E}_{\mathbf{S}_{\bm\tau}}\left[ \prod_{A \ni x} ||A||^2 \mathbb{E}_{\theta(A), \tau(A)}\left( \frac{\theta(A)(\theta(A)\tau(A) + 1)}{||A_\ell||^2(\tau(A) + 1)}\middle | S_\tau(A) \right)^{\mathbf{1}(x \in A_\ell)}\right. \\
&\qquad\qquad\qquad\qquad\left. \mathbb{E}_{\theta(A), \tau(A)}\left( \frac{(1-\theta(A))((1-\theta(A))\tau(A) + 1)}{||A_r||^2(\tau(A) + 1)}\middle | S_\tau(A) \right)^{\mathbf{1}(x \in A_r)} - \right.\\
& \qquad\qquad \left. \prod_{A \ni x} ||A||^2 \mathbb{E}_{\theta(A), \tau(A)}\left( \frac{\theta(A)^2}{||A_\ell||^2} \middle | S_\tau(A) \right)^{\mathbf{1}(x \in A_\ell)}\right. \\
&\qquad\qquad\qquad\qquad\left. \mathbb{E}_{\theta(A), \tau(A)}\left( \frac{(1-\theta(A))^2}{||A_r||^2} \middle | S_\tau(A) \right)^{\mathbf{1}(x \in A_r)}\right].
\end{align*}

\subsection*{S4: Description of simulation settings}
This section describes the generative model used in the DPM-HAPT simulation with heterogenous dispersion. The previous heterogenous dispersion model lacks degrees of freedom to create differentiated clusters with heterogenous dispersion, so here we add a fourth mixture component.

\begin{figure}[!h]
\begin{center}
\includegraphics[width=0.7\linewidth]{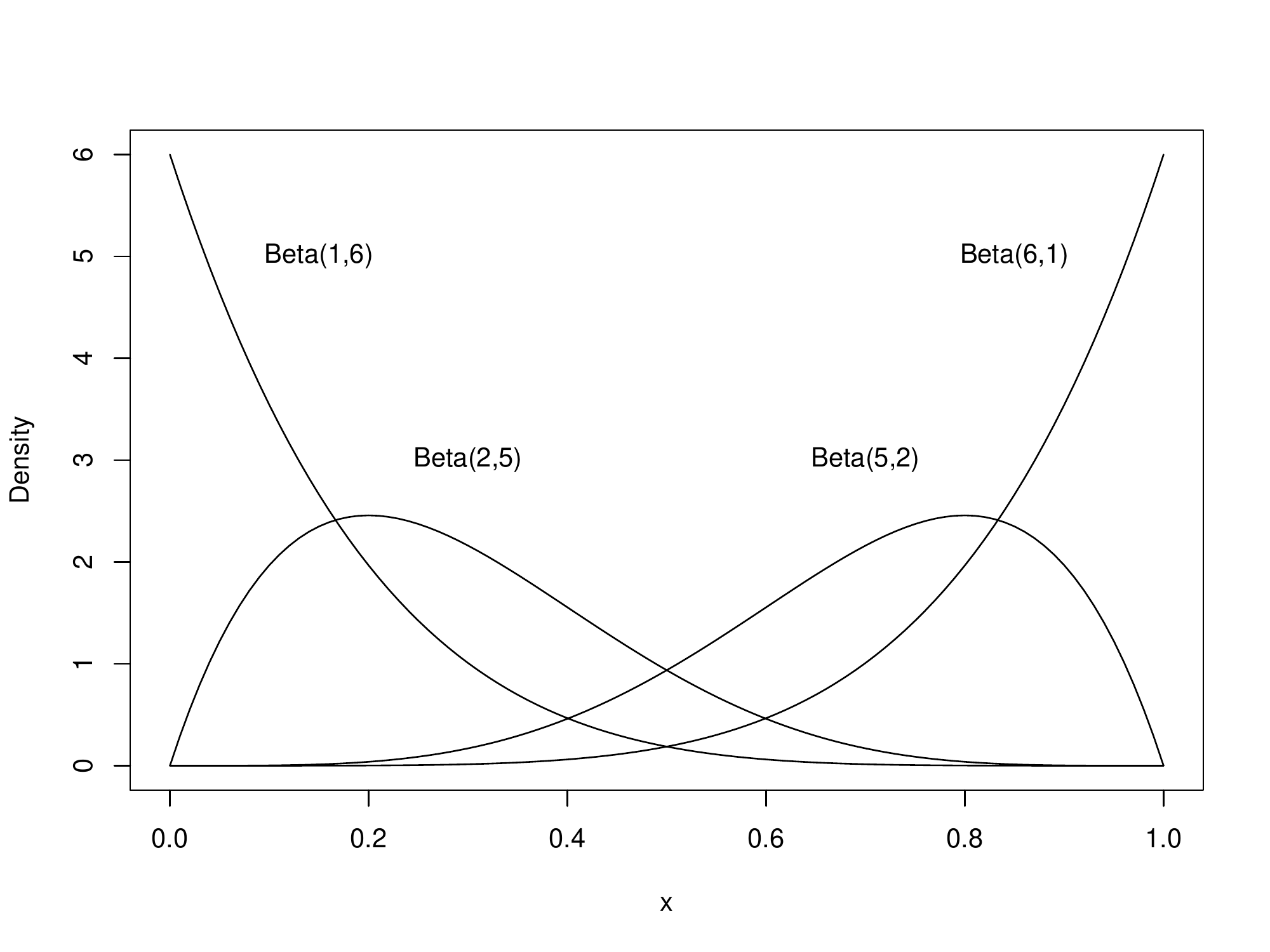}
\caption{The four mixture components used in the simulation in Sections~\ref{dispersion-process-simulations}.}
\label{mixture-components}
\end{center}
\end{figure}

Each sample is a mixture of four components, shown in Figure~\ref{mixture-components}: Beta(1,6), Beta(2,5), Beta(5,2), and Beta(6,1). The corresponding weights $w_1, w_2, w_3,$ and $w_4$ of the mixture components are drawn according to the following scheme. Let $v_1, v_2, v_3$ be Beta random variables whose parameters that differ by cluster, with $v_3$ having a much larger variance than $v_1$ and $v_2$. Then we calculate the weights as follows:
\begin{align*}
w_1 &= v_1 v_2 \\
w_2 &= v_1 (1-v_2) \\
w_3 &= (1-v_1) v_3 \\
w_4 &= (1-v_1) (1-v_3).
\end{align*}
Because $v_3$ has a much larger variance than $v_1$ and $v_2$, we end up with much more variation between densities on the right half of the space than on the left half.

For this simulation study we use
\begin{align*}
v_1 &\sim \text{Beta}(1000,1000) \\
v_2 &\sim \text{Beta}(200,200) \\
v_3 &\sim \text{Beta}(1,1).
\end{align*}

\newpage

\subsection*{S5: Sample histograms for DNase-seq data}
\label{rest-histgrams-subsection}

\begin{figure}[!h]
\begin{center}
\includegraphics[width=\linewidth]{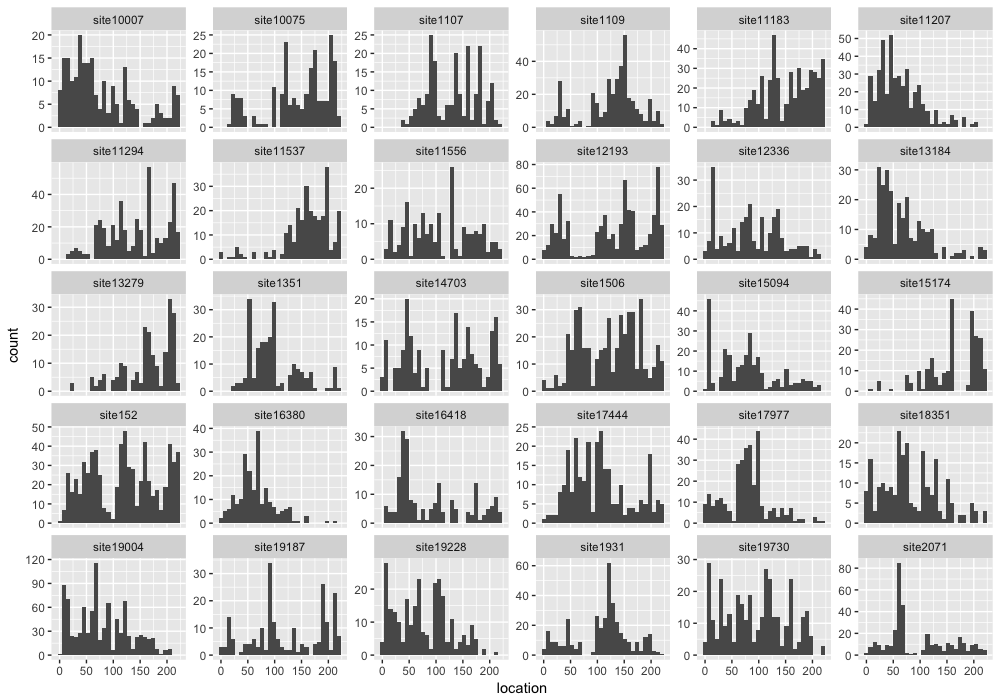}
\caption{Histograms of 30 locations from the DNase-seq data analyzed in Section~\ref{applications}.}
\label{rest-histograms}
\end{center}
\end{figure}

\end{document}